\documentclass[preprint,superscriptaddress]{revtex4}
\usepackage{graphicx}
\usepackage{lipsum}
\usepackage{float}
\usepackage{amsmath}
\usepackage{amssymb}
\usepackage{mathtools}
\usepackage{subeqnarray}

\begin{document}

\title{Statistical analysis of articulation points 
in configuration model networks}

\author{Ido Tishby}
\affiliation{Racah Institute of Physics, The Hebrew University, 
Jerusalem 91904, Israel}

\author{Ofer Biham}
\affiliation{Racah Institute of Physics, The Hebrew University, 
Jerusalem 91904, Israel}

\author{Reimer K\"uhn}
\affiliation{Department of Mathematics,
King's College London,
Strand, London WC2R 2LS, UK}

\author{Eytan Katzav} 
\affiliation{Racah Institute of Physics, The Hebrew University, 
Jerusalem 91904, Israel}

\begin{abstract}
An articulation point (AP) in a network is a node whose deletion would
split the network component on which it resides into two or more
components.
APs are vulnerable spots that play an important role in network collapse
processes, which may result from node failures, attacks or epidemics.
Therefore, the abundance and properties of APs affect the resilience
of the network to these collapse scenarios.
Here we present analytical results for the statistical properties of 
APs in configuration model networks.
In order to quantify the abundance of APs,
we calculate the probability
$P(i \in {\rm AP})$,
that a random node, $i$, in a configuration model network 
with a given degree distribution, $P(K=k)$,
is an AP.
We also obtain the conditional probability 
$P(i \in {\rm AP} | k)$
that a random
node of degree $k$ is an AP,
and find that high degree nodes are more likely to be APs than 
low degree nodes.
Using Bayes' theorem, we obtain the conditional degree distribution, 
$P(K=k | {\rm AP})$, 
over the set of APs and compare it to the overall degree distribution
$P(K=k)$.
We propose a new centrality measure based on APs:
each node can be characterized by its
articulation rank, $r$, which is the number of components
that would be added to the network upon deletion of that node.
For nodes which are not APs the articulation rank is $r=0$,
while for APs it satisfies $r \ge 1$.
We obtain a closed form analytical expression for the distribution 
of articulation ranks,
$P(R=r)$.
Configuration model networks often exhibit a coexistence between
a giant component and finite components.
While the giant component is extensive in the network size and
exhibits cycles, the finite components are non-extensive tree structures.
To examine the distinct properties of APs on the giant and on the finite components,
we calculate the probabilities presented above separately for the giant
and the finite components.
We apply these results to ensembles of configuration model networks
with degree distributions that follow 
a Poisson distribution (Erd{\H o}s-R\'enyi networks), 
an exponential distribution of the form $P(K=k) \sim e^{- \alpha k}$
and a power-law distribution 
of the form 
$P(K=k) \sim k^{- \gamma}$
(scale-free networks),
where $k \ge k_{\rm min} =1$.
The implications of these results are discussed in the context of common
attack scenarios and network dismantling processes.
\end{abstract}

\maketitle

\section{Introduction}

Network models provide a useful conceptual framework 
for the study of a large variety of systems and processes
in science, technology and society
\cite{Havlin2010,Newman2010,Estrada2011,Barrat2012}.
These models consist of nodes and edges, where the nodes
represent physical objects, while the edges represent the
interactions between them.
Unlike regular lattices in which all the nodes have the same coordination
number, network models are characterized by a degree distribution,
$P(K=k)$, $k=0,1,2,\dots$,
whose mean is denoted by 
$\langle K \rangle$.
The backbone of a network typically consists of high degree
nodes, or hubs, which connect the different branches and
maintain the integrity of the network.
In some applications,
such as communication networks,
it is crucial that the network will consist of a single connected component.
However, mathematical models also produce networks
which combine a single giant component and small isolated components,
as well as fragmented networks which consist of only small components
\cite{Bollobas2001}.

Networks are often exposed to the loss of nodes and edges, which may
severely affect their functionality. 
Such losses may occur due to inadvertent node failures,
propagation of epidemics or deliberate attacks.
Starting from a network which consists of a single connected component,
as nodes are deleted some small fragments become disconnected from the giant
component. 
As a result, the size of the giant component decreases until it 
disintegrates into many small components.
The ultimate failure, when the network fragments into small disconnected 
components was studied extensively using percolation theory
\cite{Albert2000,Cohen2000,Cohen2001a,Cohen2001b,Schneider2011,Braunstein2016,Zdeborova2016}.

\begin{figure}
\includegraphics[width=5cm]{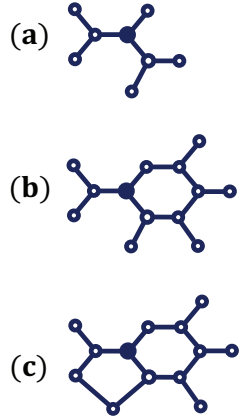} 
\caption{
(Color online)
Schematic illustration of APs and their surrounding network components:
(a) An AP (marked by a full circle) of degree $k=3$ in a tree network component.
Deletion of the AP would split the network into three separate components;
(b) An AP (full circle) of degree $k=3$, where two of its neighbors reside on a cycle.
Deletion of the AP would split the network into two separate components;
(c) Here the node marked by a full circle is not an AP because each pair of
neighbors of the marked node share a cycle. As a result, upon deletion of
the marked node all its neighbors remain on the same network component.
}
\label{fig:1}
\end{figure}

A major factor in the sensitivity of networks to node deletion processes is
the fact that the deletion of a single node may separate a whole fragment
from the giant component. This fragmentation process greatly accelerates
the disintegration of the network.
In each network, one can identify the nodes whose deletion would break
the component on which they reside into two or more components
\cite{Hopcroft1973,Gibbons1985,Tian2017}.
Such nodes are called articulation points (APs)
or cut vertices.
In Fig. \ref{fig:1}(a) we present a
schematic illustration of an AP (marked by a full circle) 
of degree $k=3$ in a tree network component.
Deletion of the AP would split the network into three separate components.
In Fig. \ref{fig:1}(b) we show an AP (full circle) of degree $k=3$, 
where two of its neighbors reside on a cycle.
Deletion of the AP would split the network into two separate components.
The node marked by a full circle in Fig. \ref{fig:1}(c) 
is not an AP because each pair of its
neighbors share a cycle. As a result, upon deletion of
the marked node all its neighbors remain on the same network component.

Since isolated nodes and leaf nodes cannot be APs,
in order for a node $i$ to be an AP, it must be of degree $k \ge 2$.
Upon deletion of node $i$ of degree $k$, in order for its $k$ neighbors
to remain on the same connected component, each pair of neighbors
must be connected to each other via at least one path in the
reduced network from which $i$ was removed. In case that upon deletion of $i$
there is at least one pair of neighbors of $i$ which are not connected to each
other, these nodes will end up on different connected components,
implying that $i$ is an AP.
In tree networks each pair of nodes are connected by a single path,
which means that any node of degree $k \ge 2$ is an AP.
As the network becomes denser and the number of cycles 
increases, the abundance of APs tends to decrease.
Since the cycles connecting neighbors of a node $i$ may be very long,
the determination of whether $i$ is an AP cannot be done locally. 
It requires access to the complete structure of the whole network
in order to identify the cycles which connect all the pairs of neighbors of $i$
\cite{Marinari2004,Marinari2006a,Marinari2006b,Marinari2007,Kerrebroeck2008}.
The statistical properties of cycles in a network are closely related to those of paths
connecting random pairs of nodes
\cite{Dorogovtsev2003,Blondel2007,Hofstad2008,Esker2008,Katzav2015,Nitzan2016,Katzav2018,Bonneau2017,Steinbock2017,Schieber2017}.
More specifically, on tree networks the shortest path is the only path connecting any
pair of nodes and thus there are no cycles.
In other networks, the distribution of cycle lengths on which a random node
resides can be obtained from the distribution of path lengths between pairs of
neighbors of a random node, $i$, in the reduced network from which $i$ is removed.
Network components which do not include any APs are called biconnected or 2-connected
components. In order to split such network components to two or more parts one
needs to simultaneously delete at least two nodes.
In such network components each pair of nodes is connected by at least two 
disjoint paths
\cite{Menger1927}.

Apart from the loss of nodes, networks are often exposed to the loss of
connections between nodes, which can be modeled by edge deletion.
In some cases the deletion of a single edge would break the network 
component on which it resides into two separate components.
Such edges are called bridges or cut-edges
\cite{Tarjan1974}, and in many ways are analogous to the
articulation points considered in this paper. 
In fact, any edge which does not reside on even a single cycle is a bridge.
Thus, in network components which exhibit a tree structure all the edges
are bridges.
A bridge, like any other edge in a network provides a connection between 
two nodes. In case that a pair of nodes, $i$ and $j$, are connected by a bridge,
each one of them must be either an AP or a leaf node. In case that the bridge
resides on the giant component, at least one of the two nodes at its ends
must be an AP.
Similarly, each AP is connected to at least one bridge.

The functionality of most networks relies on the integrity of their giant components.
Therefore, it is particularly important to study the properties of APs which reside on the 
giant component. These APs are vulnerable spots in the structure of a network,
because the deletion of a single AP may detach an entire branch 
or several branches from the giant component.
This vulnerability is exploited in network attack strategies, which target
existing APs and generate new APs via decycling processes 
\cite{Tian2017}.
While APs make the network vulnerable to attacks, they are advantageous
in fighting epidemics. 
In particular, the vaccination of APs prevents the spreading of epidemics
between the network components connected by these APs.
Similarly, in communication networks the party in possession of an AP
may control, screen, block or alter the communication between the network components
connected by this AP.
APs are instrumental in the design of efficient algorithms for 
approximate solutions of difficult computational
problems on networks, such as the vertex cover problem
\cite{Patel2014}. They can also be used to simplify the calculation
of determinants of sparse matrices which represent networks that include APs
\cite{Maybee1989}.

In this paper we present analytical results for the 
statistical properties of APs in 
Erd{\H o}s-R\'enyi (ER) networks 
\cite{Erdos1959,Erdos1960,Erdos1961}
and configuration model networks with various degree distributions
\cite{Molloy1995,Molloy1998,Newman2001,Fronczak2004}.
We obtain the probability
that a random node in a configuration model network 
with a given degree distribution $P(K=k)$
is an AP. 
We also calculate the conditional probability 
$P(i \in {\rm AP} | k)$
that a random node of a given degree $k$ is an AP and the
degree distribution 
$P(K=k | {\rm AP})$,
conditioned on the APs.
The above discussion motivates the introduction of a new
AP-based centrality measure:
when an AP is deleted the component on which it resides breaks into
two or more components. We denote the number of components, $r$,
which are added to the network upon deletion of a given node, $i$, as the 
articulation rank of this node. 
The articulation rank of a node which is not an AP is $r=0$, while the
articulation ranks of APs satisfy $r \ge 1$.
We obtain analytical results for the distribution of articulation ranks,
$P(R=r)$.

The paper is organized as follows.
In Sec. II we present the configuration model.
In Sec. III we describe relevant properties of 
the giant component and the finite components of configuration model networks.
In Sec. IV we discuss the main properties of articulation points and
present efficient methods for their detection.
In Sec. V we present analytical results for the probability that a random node in a configuration
model network is an articulation point.
In Sec. VI we present analytical results for the degree distribution of 
articulation points. 
In Sec. VII we calculate the distribution of ranks of articulation points.
In Sec. VIII we study the special properties of APs with degree $k=2$.
In Sec. IX we apply these results to configuration model networks with
Poisson degree distributions (ER networks), exponential degree distributions
and power-law degree distributions (scale-free networks).
The results are discussed in Sec. X and summarized in Sec. XI.
In Appendix A we present some useful properties of the generating
functions which are utilized in the analysis.
In Appendix B we use these inequalities to show that the mean degree
of the APs that reside on the giant component is larger than the mean
degree of all the nodes of the giant component.
In Appendix C we use the configuration model network with a ternary
degree distribution to systematically explore the abundance of APs in
the giant component and in the finite components under different 
conditions.

\section{The configuration model}

The configuration model is an ensemble of uncorrelated random networks
which follow a pre-defined degree distribution, $P(K=k)$.
In analytical studies one often considers the asymptotic case in which
the network size is infinite. 
In numerical simulations, the network size, $N$, is finite.
In many cases one bounds the degree distribution from above and below
such that $k_{\rm min} \le k \le k_{\rm max}$.
For example, using $k_{\rm min}=1$
eliminates the possibility of isolated nodes, 
while $k_{\rm min}=2$ also eliminates the leaf nodes.
Controlling the upper bound is particularly important in the case of degree
distributions which exhibit fat tails, such as power-law degree distributions.

The configuration model ensemble is a maximum entropy ensemble
under the condition that the degree distribution, $P(K=k)$, is imposed
\cite{Newman2001,Newman2010}.
Here we focus on the case of undirected networks, in 
which all the edges are bidirectional.
In each network instance from an ensemble of
configuration model networks of $N$ nodes
with a given degree distribution $P(K=k)$, one draws
the degrees of all the $N$ nodes independently from 
$P(K=k)$,
producing the degree sequence
$k_1,k_2,\dots,k_N$.

For the computer simulations presented below, we draw 
random network instances from an ensemble of configuration model
networks of $N$ nodes which follow a given degree distribution, $P(K=k)$.
For each network instance 
we generate a degree sequence of the form
$k_1, k_2,\dots,k_N$,
as described above.
For the construction process, it is convenient to order the degree
sequence in the form
$k_1 \ge k_2 \ge \dots, k_N$.
It turns out that not every possible degree sequence is graphic,
namely admissible as a degree sequence of at least one network instance.
Therefore, before trying to construct a network with a given
degree sequence, one should first confirm
the graphicality of the degree sequence.
To be graphic, a degree sequence must satisfy two conditions.
The first condition is that the sum of the degrees is an even number,
namely
$\sum\limits_{i=1}^N k_i = 2 L$,
where $L$ is an integer which represents
the number of edges in the network.
The second condition is expressed by the Erd{\H o}s-Gallai theorem,
which states that an ordered sequence of the form
$k_1 \ge k_2 \ge \dots \ge k_N$
is graphic if and only if the condition
\cite{Erdos1960b,Choudum1986}

\begin{equation}
\sum_{i=1}^n k_i \le n(n-1) + \sum_{i=n+1}^N \min (k_i,n)
\label{eq:EG}
\end{equation}

\noindent
holds for all values of $n$ in the range
$1 \le n \le N-1$.

A convenient way to construct a configuration model network 
is to prepare the $N$ nodes such that each node, $i$, is 
connected to $k_i$ half edges or stubs
\cite{Newman2010}.
Pairs of half edges 
from different nodes
are then chosen randomly
and are connected to each other in order
to form the network. 
The result is a network with the desired degree sequence and
no correlations.
Note that towards the end of the construction
the process may get stuck.
This may happen in case that the only remaining pairs of stubs
belong to the same node or to nodes which are already connected to each other.
In such cases one may perform some random reconnections 
in order to enable completion of the construction.

\section{The giant component and the finite components}

Configuration model networks often consist of multiple connected 
components. In some cases the size of the largest component
scales linearly with the network size, $N$. In such cases, the largest
component is called the giant component.
All the other components are finite, non-extensive, components
which exhibit tree structures with no cycles. 
The size of the giant component is determined by the 
degree distribution, $P(K=k)$.
Some families of degree distributions can be parametrized such
that the parameter space is separated into two regimes,
the dilute network regime in which there is no giant component
and the dense network regime in which there is a giant component.
On the boundary between these two regimes there is a percolation transition
\cite{Havlin2010}.

Consider a configuration model network 
of $N$ nodes with a given degree distribution, $P(K=k)$.
To obtain the the probability, $g$, that a random node
in the network belongs to the giant component,
one needs to first calculate the probability $\tilde g$,
that a random neighbor of a random node, $i$,
belongs to the giant component of the reduced network, 
which does not include the node $i$.
The probability $\tilde g$ is determined by
\cite{Havlin2010}

\begin{equation}
1 - {\tilde g} = G_1(1 - {\tilde g}),
\label{eq:tg}
\end{equation}

\noindent
where

\begin{equation}
G_1(x) = \sum_{k=1}^{\infty}   x^{k-1}   {\widetilde P}(K=k) 
\label{eq:G1}
\end{equation}

\noindent
is the generating function of the distribution
${\widetilde P}(K=k)$, 
which is the degree distribution of nodes which are sampled as 
random neighbors of random nodes.
It is given by

\begin{equation}
{\widetilde P}(K=k) = \frac{k}{\langle K \rangle} P(K=k),
\label{eq:tilde}
\end{equation}

\noindent
where

\begin{equation}
\langle K \rangle = \sum_{k=0}^{\infty} k P(K=k)
\end{equation}

\noindent
is the mean degree of the nodes in the network.
Using $\tilde g$, one can then obtain the probability $g$ from the equation

\begin{equation}
g = 1 - G_0(1 - {\tilde g}),
\label{eq:g}
\end{equation}

\noindent
where

\begin{equation}
G_0(x) = \sum_{k=0}^{\infty}   x^{k}   P(K=k) 
\label{eq:G0}
\end{equation}

\noindent
is the generating function of the distribution $P(K=k)$.

From the definitions of $G_0(x)$ and $G_1(x)$ in Eqs.
(\ref{eq:G0})
and 
(\ref{eq:G1}),
respectively,
we find that
$G_0(1)=1$
and
$G_1(1)=1$.
This 
means that $x=1$ is a fixed point for both generating functions.
Therefore, $g=\tilde g=0$ is a solution of Eq. (\ref{eq:tg}).
This solution corresponds to the case of subcritical networks,
in which there is no giant component.
In some networks there are no isolated nodes (of degree $k=0$)
and no leaf nodes (of degree $k=1$). 
In such networks
$P(K=k) > 0$ only for $k \ge 2$. 
For these networks
we find that
$G_0(0) = 0$ 
and
$G_1(0)=0$.
This implies that in such networks both $x=0$ and $x=1$ are fixed points
of both $G_0(x)$ and $G_1(x)$.
The coexistence of a giant component and finite components appears
for degree distributions that support a non-trivial solution of Eq. (\ref{eq:tg}),
in which $0 < \tilde g < 1$.

The probability that a random node
resides on the giant component (GC) is
$P(i \in {\rm GC}) = g$,
and the probability that it resides on one of the finite components (FC) is
$P(i \in {\rm FC}) = 1 - g$.
Similarly, the probabilities that a random neighbor of a random node resides
on the giant 
component is
$\widetilde P(i \in {\rm GC}) = \tilde g$
and the probability that it resides on one of the finite components is
$\widetilde P(i \in {\rm FC}) = 1 - \tilde g$.
A node, $i$, of degree $k$ resides on the giant component if at least one of its $k$ neighbors
resides on the giant component of the reduced network from which $i$ is removed.
Therefore,

\begin{equation}
P(i \in {\rm GC} | k)= 1 - (1 - \tilde g)^k,
\label{eq:L1k} 
\end{equation}

\noindent
and

\begin{equation}
P(i \in {\rm FC} | k) =  (1 - \tilde g)^k.
\label{eq:L0k}
\end{equation}

The micro-structure of the giant component of configuration model
networks was recently studied
\cite{Tishby2018}.
It was shown that the degree distribution,
conditioned on the giant component, is given by

\begin{equation}
P(K=k | {\rm GC}) =
\frac{ 1- (1-\tilde g)^k}{g} P(K=k),
\label{eq:kL1}
\end{equation}

\noindent
while the degree distribution,
conditioned on the finite components,
is given by

\begin{equation}
P(K=k | {\rm FC}) =
\frac{ (1-\tilde g)^k}{1 - g} P(K=k).
\label{eq:kL0}
\end{equation}

\noindent
The mean degree of the giant component is 

\begin{equation}
\mathbb{E}[K | {\rm GC}] =
\frac{1-(1-\tilde g)^2}{g} 
\langle K \rangle,
\label{eq:EkL1}
\end{equation}

\noindent
or

\begin{equation}
\mathbb{E}[K | {\rm GC}] =
\frac{\tilde g (2-\tilde g)}{g} 
\langle K \rangle,
\label{eq:EkL1b}
\end{equation}

\noindent
while the mean degree on the finite components is

\begin{equation}
\mathbb{E}[K | {\rm FC}] =
\frac{(1-\tilde g)^2}{1-g} 
\langle K \rangle,
\label{eq:EkL0}
\end{equation}

\noindent
Using the inequality
$(1-\tilde g)^2/(1-g) < 1$
(Appendix A),
we find that, as expected, 
$\mathbb{E}[K| {\rm GC}] > \langle K \rangle$
and
$\mathbb{E}[K| {\rm FC}] < \langle K \rangle$.

\section{Articulation points and their detection}

An articulation point in a network is a node, $i$, whose deletion would break
the network component on which it resides into two or more components.
Each one of these components must include at least one neighbor of $i$.
Therefore, the degree of an AP must satisfy the condition $k \ge 2$.
To determine whether a given node,
$i$, of degree $k \ge 2$, is an AP, we first delete it and mark its $k$ neighbors. 
We randomly choose one
of these neighbors and label all the nodes which belong to the connected component
on which it resides. This can be done by using either the breadth first search (BFS) or
the depth first search (DFS) algorithms. If all the other $k-1$ neighbors of $i$ belong to
this connected component, then node $i$ is not an AP. 
Alternatively, if at least one of these neighbors does not belong to this component,
then node $i$ is an AP. 

The above approach is useful in order to determine whether a given node is an AP.
However, it is not an efficient approach for finding all the APs in the network.
This is due to the fact that in this approach one needs to repeat the cluster labeling
procedure separately for each node in the network. 
It turns out that there is a more efficient algorithm, which enables one to detect
all the APs in a network using a single DFS run,
with respect to a randomly selected reference node
\cite{Hopcroft1973,Gibbons1985}.
Below we demonstrate this algorithm for the giant component of a
configuration model network.
In order to find all the APs in a configuration model network, we choose a 
random node, $i$, on the giant component. We then run a DFS around the
root node $i$ over the whole giant component. Using this approach we
essentially determine the shell structure around $i$. The first shell
consists of the neighbors of $i$. The second shell consist of 
nodes which are at distance $2$ from $i$, while the $\ell$th
shell consists of nodes which are at distance $\ell$ from $i$.
For each node in the $\ell$th shell, we maintain
a record of its neighbors in the $(\ell-1)$th, 
$\ell$th and $(\ell+1)$th shells.

Consider a node, $j$, which resides in the $\ell$th shell and has $k'$
neighbors in the $(\ell+1)$th shell, denoted by $j_m$, $m=1,2,\dots,k'$.
To determine whether the node $j$ is an AP, we first delete it from the network.
We then check for each one of its $k'$ neighbors in the $(\ell+1)$th shell
whether it has a path to the root node, $i$, in the reduced network from
which $j$ was removed. If all these $k'$ neighbors have such paths to node
$i$ then the node $j$ is not an AP. However, if at least one of these $k'$ 
neighbors does not have 
such path to $i$ then the node $j$ is an AP. 
This procedure should be repeated for all the nodes, $j$, in the network.
The algorithm presented above provides all the APs which reside in the 
giant component. In the finite tree components there is no need to apply
the algorithm since all the nodes of degrees $k \ge 2$ are APs.

\section{The probability that a random node is an articulation point}

In this section we derive a closed form analytical expression for the probability 
$P(i \in {\rm AP})$
that a random node, $i$,
is an AP.
To this end, we first consider the conditional probability,
$P(i \in {\rm AP} | k)$,
that a random node of a given degree, $k$,
is an AP.
This probability can be expressed by
$P(i \in {\rm AP} | k ) = 1 - P(i \notin {\rm AP} | k)$,
where $P(i \notin {\rm AP} | k)$
is the probability that a random node of degree $k$ is
not an AP.
Clearly, nodes of degree $k=0$ or $1$ cannot be APs.
This is due to the fact that nodes of degree $k=0$ are
isolated from the rest of the network, while nodes of 
degree $k=1$ are leaf nodes whose deletion does not
affect the connectivity of the rest of the network. 
Therefore,

\begin{equation}
P(i \in {\rm AP} | 0) = P(i \in {\rm AP} | 1) = 0.
\end{equation}

\noindent
In order that a node $i$ of degree $k \ge 2$ will not be an AP,
all its $k$ neighbors must reside on the giant component of
the reduced network from which $i$ is removed.
This occurs with probability
$P(i \notin {\rm AP} | k) =  {\tilde g}^k$.
Therefore, 

\begin{equation}
P(i \in {\rm AP} | k) =  (1 - {\tilde g}^k) \theta(k-2),
\label{eq:APk}
\end{equation}

\noindent
where $\theta(k)$ is the Heaviside step function,
which satisfies $\theta(k)=1$ for $k \ge 0$ and
$\theta(k)=0$ for $k<0$.
Thus, the probability that a random node of unspecified degree
is an AP is given by

\begin{equation}
P(i \in {\rm AP})
= \sum_{k=2}^{\infty}  \left( 1 - {\tilde g}^k \right) P(K=k) .
\label{eq:AP1}
\end{equation}

\noindent
This probability can also be expressed in the form

\begin{equation}
P(i \in AP) = 1  - G_0(\tilde g) - (1 - \tilde g) P(K=1),
\label{eq:AP2}
\end{equation}

\noindent
where $G_0(x)$ is given by Eq. (\ref{eq:G0}).
Note that these results are based on the assumption that the
probabilities that different neighbors of a random node, $i$, 
reside on the giant component are independent.
This assumption is expected to hold in ensembles of uncorrelated networks,
such as the configuration model networks.

An important distinction in configuration model networks is
between the properties of nodes which reside on the giant
component and those which reside on the finite components.
In particular, the probability $P(i \in {\rm AP})$ can be
expressed as a sum of the contributions of the giant and finite components:

\begin{equation}
P(i \in {\rm AP}) = 
P(i \in {\rm AP} | {\rm GC}) P( i \in {\rm GC})
+
P(i \in {\rm AP} | {\rm FC}) P(i \in {\rm FC}),
\label{eq:AP01}
\end{equation}

\noindent
For nodes which reside on the finite components

\begin{equation}
P(i \in {\rm AP} | {\rm FC}) =
1 - P(K=0 | {\rm FC}) - P(K=1 | {\rm FC}).
\label{eq:APL0p}
\end{equation}

\noindent
Using Eq. (\ref{eq:kL0}) we find that

\begin{equation}
P(K=0 | {\rm FC}) = \frac{P(K=0)}{1-g},
\end{equation}

\noindent
and

\begin{equation}
P(K=1 | {\rm FC}) = \frac{1-\tilde g}{1-g} P(K=1).
\end{equation}

\noindent
Therefore, the probability that a random node that resides on one
of the finite components is an AP is given by

\begin{equation}
P(i \in {\rm AP} | {\rm FC}) = 
1 -   \frac{1}{1-g} P(K=0) 
- \frac{1-\tilde g}{1-g} P(K=1).
\label{eq:APL0}
\end{equation}

\noindent
Inserting this result into Eq. (\ref{eq:AP01}), one can extract the
probability that a random node that resides on the giant component
is an AP. This probability is given by

\begin{equation}
P(i \in {\rm AP} | {\rm GC}) =
1 - \frac{1}{g}
G_0(\tilde g)
+ \frac{1}{g} P(K=0).
\label{eq:APL1}
\end{equation}

\noindent
Using Bayes' theorem,
the probability that a random AP in the network resides on the giant component
can be expressed in the form

\begin{equation}
P(i \in {\rm GC} | {\rm AP}) =
\frac{P(i \in {\rm AP} | {\rm GC}) P(i \in {\rm GC})}
{P(i \in {\rm AP})}.
\end{equation}

\noindent
Inserting $P(i \in {\rm AP}) | {\rm GC})$
from Eq. (\ref{eq:APL1}),
we obtain

\begin{equation}
P(i \in {\rm GC} | {\rm AP}) =
\frac{g  - G_0(\tilde g) + P(K=0)
}
{1  -  G_0(\tilde g) - (1 - \tilde g) P(K=1)
}.
\label{eq:L1AP}
\end{equation}

\noindent
Thus, the complementary probability, that a random AP in the network resides on one of the
finite components, is given by

\begin{equation}
P(i \in {\rm FC} | {\rm AP}) =
\frac{1 - g - P(K=0) - (1-\tilde g) P(K=1)  }
{1  -  G_0(\tilde g) - (1 - \tilde g) P(K=1)
}.
\label{eq:L0AP}
\end{equation}

\noindent
The probability that a random node of degree $k$ is an AP,
given by Eq. (\ref{eq:APk}), 
can be expressed as a sum
of the contributions of the giant and finite components,
in the form

\begin{eqnarray}
P(i \in {\rm AP} | k) &=&
P(i \in {\rm AP} |{\rm GC}, k) P(i \in {\rm GC}| k)
\nonumber \\ 
&+&
P(i \in {\rm AP} |{\rm FC}, k) P(i \in {\rm FC}| k).
\label{eq:APk01}
\end{eqnarray}

\noindent
For nodes which reside on the finite components

\begin{equation}
P(i \in {\rm AP} | {\rm FC}, k) = \theta(k-2).
\label{eq:APL0k}
\end{equation}

\noindent
Inserting Eqs. 
(\ref{eq:L1k}), (\ref{eq:L0k}),
(\ref{eq:APk}) and (\ref{eq:APL0k}) in 
Eq. (\ref{eq:APk01})
we obtain

\begin{equation}
P(i \in {\rm AP} | {\rm GC}, k) =
\left[ 1 - \frac{\tilde g^k}{1 - (1-\tilde g)^k} \right]
\theta(k-2).
\label{eq:APL1k}
\end{equation}

\noindent
The probability that a random AP of degree $k$ resides on the giant component
is given by

\begin{equation}
P(i \in {\rm GC}| {\rm AP}, k) =
\frac{P(i \in {\rm AP} | {\rm GC}, k) P(i \in {\rm GC}| k)}
{P(i \in {\rm AP} | k)}.
\end{equation}

\noindent
Inserting 
$P(i \in {\rm AP} | {\rm GC}, k)$
from Eq. (\ref{eq:APL1k})
we obtain

\begin{equation}
P(i \in {\rm GC} | {\rm AP}, k) =
1 - \frac{ (1 - \tilde g)^k }{ 1 - \tilde g^k }, \ \ \ k \ge 2.
\label{eq:L1APk}
\end{equation}

\noindent
The probability that a random AP of degree $k$ resides on 
one of the finite components
is given by

\begin{equation}
P(i \in {\rm FC} | {\rm AP}, k) =
\frac{P(i \in {\rm AP} | {\rm FC}, k) P(i \in {\rm FC}| k)}
{P(i \in {\rm AP} | k)}.
\label{eq:L0APkp}
\end{equation}

\noindent
Inserting 
$P(i \in {\rm AP} | {\rm GC}, k)$
from Eq. (\ref{eq:APL1k})
into Eq. (\ref{eq:L0APkp}),
we obtain

\begin{equation}
P(i \in {\rm FC} | {\rm AP}, k) =
\frac{   (1-\tilde g)^k   }
{1 - \tilde g^k}, \ \ \ k \ge 2.
\label{eq:L0APk}
\end{equation}

\noindent
Comparing this result to the expression for $P(i \in {\rm FC}| k)$, given by 
Eq. (\ref{eq:L0k}),
it is found that
for $k \ge 2$ and $0 < \tilde g < 1$,
$P(i \in {\rm FC}|{\rm AP}, k) > P(i \in {\rm FC}| k)$.
This result means that an AP of degree $k \ge 2$ is more likely to
reside on one of the finite components than a random node of the same degree.

In the limit of $\tilde g \rightarrow 1$,
the giant component encompasses the whole network.
In this case, for $k \ge 2$ the probability
$P(i \in {\rm GC}|{\rm AP}, k)=1$ 
and
$P(i \in {\rm FC}|{\rm AP}, k)=0$. 
In the opposite limit of $\tilde g \rightarrow 0$,
the network consists of finite components and thus
$P(i \in {\rm GC}|{\rm AP}, k)=0$ 
and
$P(i \in {\rm FC}|{\rm AP}, k)=1$. 
For $0 < \tilde g < 1$, the probability
$P(i \in {\rm GC}|{\rm AP}, k)$
is a monotonically increasing function of $k$,
while
$P(i \in {\rm FC}|{\rm AP}, k)$
is a monotonically decreasing function of $k$.
Thus, most APs of high degrees reside on the giant component
while the APs which reside on the finite components tend to be
of lower degrees.

\section{The degree distribution of the articulation points}

Using Bayes' theorem 
one can express the degree distribution 
of the APs in the form

\begin{equation}
P(K=k |  {\rm AP}) = \frac{ P(i \in {\rm AP} | k) }{ P(i \in {\rm AP}) } P(K=k).
\end{equation}

\noindent
Inserting $P(i \in {\rm AP})$ 
from Eq. (\ref{eq:AP2})
and
$P(i \in {\rm AP}| k)$ 
from Eq. (\ref{eq:APk}).
we obtain

\begin{equation}
P(K=k | {\rm AP}) = 
\frac{ (1 - \tilde g^k) \theta(k-2) }
{ 1  - 
G_0(\tilde g) - (1 - \tilde g) P(K=1)
} P(K=k).
\label{eq:kAP}
\end{equation}

\noindent
The mean degree of the APs in the network is given by

\begin{equation}
\mathbb{E}[K | {\rm AP}] = \sum_{k=2}^{\infty} k P(K=k |{\rm AP}).
\label{eq:EkAPp}
\end{equation}

\noindent
Carrying out the summation we obtain

\begin{equation}
\mathbb{E}[K |{\rm AP}] = 
\frac{   \left[ 1   -  \tilde g  G_1(\tilde g) \right] }
{ 1  -  G_0(\tilde g) - (1 - \tilde g) P(K=1)}
\langle K \rangle
-
\frac{ (1 - \tilde g) P(K=1) }
{ 1  -  G_0(\tilde g) - (1 - \tilde g) P(K=1) },
\label{eq:EkAP}
\end{equation}

\noindent
where $G_1(\tilde g)$ is given by Eq. (\ref{eq:G1}).

The degree distribution $P(K=k | {\rm AP})$, given by 
Eq. (\ref{eq:kAP}),
can be expressed as a weighted sum of the degree distributions 
of the giant and finite components, in the form

\begin{eqnarray}
P(K=k | {\rm AP}) &=& 
P(K=k | {\rm AP},{\rm GC}) P(i \in {\rm GC} | {\rm AP})
\nonumber \\
&+&
P(K=k | {\rm AP},{\rm FC}) P(i \in {\rm FC} | {\rm AP}).
\label{eq:PkAP01}
\end{eqnarray}

\noindent
Since on the finite components, all the nodes of degree $k \ge 2$ are
APs, the degree distribution over the APs which reside on the finite
components is given by

\begin{equation}
P(K=k | {\rm AP},{\rm FC}) = 
\frac{ \theta(k-2)  P(K=k | {\rm FC})  }{1 - P(K=0 |{\rm FC}) - P(K=1 |{\rm FC})}.
\label{eq:PkAP0}
\end{equation}

\noindent
Inserting the expression for
$P(K=k|{\rm FC})$,
given by 
Eq. (\ref{eq:kL0}),
into Eq. (\ref{eq:PkAP0})
we obtain

\begin{equation}
P(K=k | {\rm AP},{\rm FC}) =
\frac{(1 - \tilde g)^k  \theta(k-2)}
{1-g - P(K=0) - (1 - \tilde g) P(K=1)}
P(K=k).
\label{eq:kAPL0}
\end{equation}

\noindent
Inserting this result into Eq. (\ref{eq:PkAP01})
we obtain

\begin{equation}
P(K=k | {\rm AP},{\rm GC}) =
\left[
\frac{1 - \tilde g^k - (1-\tilde g)^k}
{g - G_0(\tilde g) + P(K=0)} 
\right]
\theta(k-2) P(K=k) .
\label{eq:kAPL1}
\end{equation}

\noindent
The expectation values of the degrees
of the APs that reside on the giant and finite
components are given by

\begin{equation}
\mathbb{E}[K | {\rm AP},\Lambda] 
= \sum_{k=2}^{\infty} k P(K=k | {\rm AP},\Lambda),
\end{equation}

\noindent
where $\Lambda={\rm GC}$ and ${\rm FC}$, respectively.
Carrying out the summation, we obtain

\begin{equation}
\mathbb{E}[K | {\rm AP},{\rm GC}] =
\frac{ 1 - \tilde g  G_1(\tilde g) - (1-\tilde g)^2  }
{g - G_0(\tilde g) + P(K=0)} \langle K \rangle,
\label{eq:EkAPL1}
\end{equation}

\noindent
and

\begin{equation}
\mathbb{E}[K | {\rm AP},{\rm FC}] = 
\frac{ 
(1-\tilde g)^2   \langle K \rangle - (1-\tilde g) P(K=1)   }
{ 1 - g - P(K=0) - (1 - \tilde g) P(K=1)  }.
\label{eq:EkAPL0}
\end{equation}

\noindent
In Appendix B we show that the mean degree of APs on the giant
component is larger than the mean degree of all nodes on the
giant component, namely

\begin{equation}
\mathbb{E}[K | {\rm AP},{\rm GC}] > \mathbb{E}[K | {\rm GC}].
\end{equation}

\noindent
This is consistent with the fact that high degree nodes are more
likely to be APs than low degree nodes. 
Similarly, the mean degree of APs on the finite components is larger than the mean
degree of all nodes on the finite components, namely

\begin{equation}
\mathbb{E}[K | {\rm AP},{\rm FC}] > \mathbb{E}[K | {\rm FC}].
\end{equation}

\noindent
This can be easily understood as follows.
The finite components exhibit tree structures, whose branches terminate
by leaf nodes of degree $k=1$. In some cases there are also isolated nodes
of degree $k=0$.
While isolated nodes and leaf nodes cannot be APs, on tree components
all the nodes of degree $k \ge 2$ are APs.
Therefore, the mean degree of the APs is larger than the mean degree of
all the nodes on the finite components.

\section{The distribution of articulation ranks}

The articulation rank, $r$, of a node $i$ is defined as the number of components which
are added to the network upon deletion of $i$.
A node which is not an AP has a rank of $r=0$, because its deletion
does not add any new component to the network.
A node is an AP of rank $r \ge 1$
if its deletion
breaks the network component on which
it resides into $r+1$ parts, 
thus increasing the number of components by $r$.
The deletion of an AP of articulation rank $r$ which resides on
the giant component reduces the size of the giant component
while adding $r$ finite components to the
network.
The deletion of an AP of rank $r$ which resides on one of the
finite components breaks this component into $r+1$ fragments.
The articulation rank of a node of degree $k$
may take values in the range 
$r=0,1,2,\dots,k-1$.

On the finite components all the nodes of degrees $k \ge 2$ are APs.
Since the finite components are tree networks, the rank of a node
of degree $k$ is $r=k-1$.
Therefore, the
probability that a randomly chosen node of degree $k$,
which resides on one of the finite components
is of articulation rank $r$, is given by

\begin{equation}
P(R=r | {\rm FC}, k) =
\left\{
\begin{array}{ll}
\delta_{r,0} & \ \ \  k =0,1 \\
\delta_{r, k-1} & \ \ \  k \ge 2,
\end{array}
\right.
\label{eq:rAPL0k}
\end{equation}

\noindent
where $\delta_{k,k'}$ is the Kronecker symbol.
The probability that a randomly chosen node of degree $k$
which resides on the giant component
is of rank $r$, 
is given by

\begin{equation}
P(R=r | {\rm GC}, k) = 
\frac{1}{P(i \in {\rm GC}|k)}
\binom{k}{r} ( 1 - \tilde g )^r   \tilde g^{k-r},
\label{eq:rAPL1k}
\end{equation}

\noindent
where 
$P(i \in {\rm GC}|k)$ is given by 
Eq. (\ref{eq:L1k})
and
$r=0,1,\dots,k-1$.
The product 
$( 1 - \tilde g )^r   \tilde g^{k-r}$
represents the probability that a randomly chosen set of $r$ neighbors of a random
node $i$ do not reside on the giant component of the reduced network from which
$i$ was removed, while the remaining $k-r$ neighbors do reside on the giant component.
The binomial coefficients account for the number of different possibilities to
choose a set of $r$ neighbors out of the $k$ neighbors of $i$.
Summing up Eqs. (\ref{eq:rAPL0k}) and (\ref{eq:rAPL1k}) over
all values of $k \ge r+1$, we obtain

\begin{equation}
P(R=r |  \Lambda) =
\sum_{k=r+1}^{\infty}
P(R=r |  \Lambda, k)
P(K=k |  \Lambda),
\end{equation}

\noindent
where $\Lambda={\rm FC}$ for nodes which reside on one
of the finite components and $\Lambda={\rm GC}$ for nodes
which reside on the giant component.
Carrying out the summation, we obtain

\begin{equation}
P(R=r |  {\rm FC}) =
\left\{
\begin{array}{ll}
\frac{1}{1-g} P(K=0) + \frac{1-\tilde g}{1-g} P(K=1)   & \ \  r=0 \\
\frac{(1 - \tilde g)^{r+1}  }{1-g }  P(K=r+1)   & \ \   r \ge 1.
\end{array}
\right.
\label{eq:rL0}
\end{equation}

\noindent
and

\begin{eqnarray}
P(R=r |  {\rm GC}) &=& 
\frac{  \left( 1-\tilde g \right)^r  }{g}
\sum_{k=r+1}^{\infty}
\binom{k}{r} 
\tilde g^{k-r}
P(K=k).
\label{eq:rL1}
\end{eqnarray}

\noindent
The mean articulation rank of the nodes which reside on the finite components is

\begin{equation}
\mathbb{E}[R| {\rm FC}] =
\frac{(1-\tilde g)^2}{1-g} \langle K \rangle - 1
+ \frac{P\left(K=0\right)}{1-g},
\label{eq:ErL0}
\end{equation}

\noindent
while the mean articulation rank of the nodes which reside on the giant 
component is

\begin{equation}
\mathbb{E}[R| {\rm GC}] =
\frac{\tilde g (1-\tilde g)}{g} \langle K \rangle.
\label{eq:ErL1}
\end{equation}

\noindent
Interestingly, using the results shown in Appendix A 
[Eq. (\ref{eq:ggt0})],
it is found that at the percolation threshold, $c=c_0$,
the mean articulation rank on the 
giant component satisfies $\mathbb{E}[R| {\rm GC}]=1$.
The distribution of articulation ranks over the whole network can be
expressed in the form

\begin{equation}
P(R=r) = P(R=r|{\rm GC})P({\rm GC}) + P(R=r|{\rm FC})P(i \in {\rm FC}),
\label{eq:Erall}
\end{equation}

\noindent
where $P(R=r| {\rm GC})$ is given by
Eq. (\ref{eq:rL1}) and $P(R=r| {\rm FC})$ is given
by Eq. (\ref{eq:rL0}).
More explicitly, it takes the form

\begin{equation}
P(R=r) =
\left\{
\begin{array}{ll}
P(K=0) + (1-\tilde g) P(K=1) + \sum\limits_{k=1}^{\infty} \tilde g^k P(K=k) & \ \ \  r = 0 \\
(1-\tilde g)^{r+1} P(K=r+1) + 
\left( 1-\tilde g \right)^r
\sum\limits_{k=r+1}^{\infty}
\binom{k}{r} 
\tilde g^{k-r}
P(K=k)  & \ \ \  r \ge 1.
\end{array}
\right.
\label{eq:rAPall}
\end{equation}

\noindent
The mean articulation rank of the whole network is

\begin{equation}
\langle R \rangle = 
(1 - \tilde g) \langle K \rangle - (1-g) + P(K=0).
\end{equation}

\section{Properties of articulation points of degree $k=2$}

Consider an AP of degree $k=2$ that resides on the GC.
One can distinguish between two types of such APs,
according to the structural properties of the tree
that is detached from the GC upon their deletion.
In one type of $k=2$ APs, referred to as {\it stubs},
the detached tree does not include any node of 
degree $k > 2$.
Hence, stubs do not bridge between the GC and
any branching trees.
The other type of $k=2$ APs,
referred to as {\it tubes},
connect the GC with a branching tree
that includes at least one node of degree $k \ge 3$.
Below we derive a closed form expression for the
probability $P({\rm Stub})$ that a random $k=2$ AP 
that resides on the GC is a stub.
A key observation is that $P({\rm Stub})$ 
is the probability that the detached tree is
a chain of any length, that consists of nodes
of degree $k=2$ and a single node of degree $k=1$
at the end.
The probability that the detached tree is a chain
of length $\ell$ is denoted by $P({\rm Stub},\ell)$.
It can be expressed in the form

\begin{equation}
P({\rm Stub},\ell) = \widetilde P(K=2|{\rm FC})^{\ell-1} P(K=1|{\rm FC}),
\end{equation}

\noindent
where
 
\begin{equation}
\widetilde P(K=k|{\rm FC}) = \frac{k}{\mathbb{E}[K|{\rm FC}]} P(K=k|{\rm FC})
\label{eq:kFC}
\end{equation}
  
\noindent
is the probability that a random neighbor of a random node
on a finite component
is of degree $k$.
The conditioning on a finite component is due to the fact
that once the AP is deleted the detached tree follows the 
statistical properties of the finite components.
Since a stub can be of any length the probability 
$P({\rm Stub})$ is given by

\begin{equation}
P({\rm Stub}) = \sum_{\ell=1}^{\infty}
P({\rm Stub},\ell).
\end{equation}

\noindent
Carrying out the summation on the right hand side 
we obtain

\begin{equation}
P({\rm Stub}) =
\frac{ \widetilde P(K=1|{\rm FC}) }{1 - \widetilde P(K=2|{\rm FC})}.
\end{equation}

\noindent
Using Eqs. (\ref{eq:kFC}), (\ref{eq:kL0}) and (\ref{eq:EkL0}),
we can express $P({\rm Stub})$ in the form

\begin{equation}
P({\rm Stub}) = \frac{P(K=1)}{(1-\tilde g)[ \langle K \rangle - 2 P(K=2)]}.
\label{eq:Stub1}
\end{equation}

\noindent
The probability that a random $k=2$ AP is a tube is thus
$P({\rm Tube}) = 1 - P({\rm Stub})$.

\section{Applications to specific network models}

Here we apply the approach presented above 
to three examples of configuration model networks, 
with a Poisson degree distribution (ER networks),
an exponential degree distribution and
a power-law degree distribution (scale-free networks).

\subsection{Erd{\H o}s-R\'enyi networks}

The ER network is the simplest kind of
a random network, 
and a special case of the configuration model.
It is a maximum entropy network under the condition
in which only the mean degree,
$\langle K \rangle = c$, 
is constrained.
ER networks can be constructed by independently connecting
each pair of nodes with probability
$p = {c}/{(N-1)}$.
In the asymptotic limit the resulting degree distribution follows a
Poisson distribution of the form

\begin{equation}
P(K=k) = \frac{e^{-c} c^k}{k!}.
\label{eq:poisson}
\end{equation}

\noindent
Asymptotic ER
networks exhibit a percolation transition at $c=1$, such that for
$c<1$
the network consists only of finite components,
which exhibit tree topologies.
The degree distribution and the distribution of shortest path lengths
on the finite components of subcritical ER networks were studied in Ref. 
\cite{Katzav2018}.
For  
$c>1$
a giant component emerges, coexisting with the finite components. 
At a higher value of the connectivity, namely at 
$c = \ln N$, 
there is a second transition, above which
the giant component encompasses 
the entire network. 

ER networks exhibit a special property, 
resulting from the Poisson degree distribution
[Eq. (\ref{eq:poisson})], 
which satisfies
$\widetilde P(K=k) = P(K=k-1)$,
where ${\widetilde P}(K=k)$ is given by Eq. (\ref{eq:tilde}).
This implies that for the Poisson distribution, 
the two generating functions 
are identical, namely $G_1(x)=G_0(x)$
\cite{Katzav2015,Nitzan2016}.
Using Eqs. (\ref{eq:tg}) and (\ref{eq:g}) we obtain that for ER networks
${\tilde g} = g$. 
Thus, for ER networks, Eq. (\ref{eq:APk}) can be replaced by

\begin{equation}
P(i \in {\rm AP} | k) = (1 - g^k) \theta(k-2).
\label{eq:APkER}
\end{equation}

\noindent
Carrying out the summation in Eq. (\ref{eq:G1})
with $P(K=k)$ given by Eq. (\ref{eq:poisson}), 
one obtains
$G_0(x)=G_1(x)=e^{-(1-x)c}$.
Inserting this result in Eq. (\ref{eq:g}),
it is found that $g$ satisfies the equation
$1-g =  e^{-gc}$
\cite{Bollobas2001}.
Solving for the probability $g$ as a function of the mean degree, $c$, 
one obtains

\begin{equation}
g = 1 + \frac{W(-c e^{-c})}{c},
\label{eq:g(c)}
\end{equation}

\noindent
where $W(x)$ is the Lambert $W$ function
\cite{Olver2010}.
Inserting $g$ from 
Eq. (\ref{eq:g(c)}) 
into Eq. (\ref{eq:APkER}),
we obtain

\begin{equation}
P(i \in {\rm AP} | k) = \left\{ 1 - \left[ 1 + \frac{W(-c e^{-c})}{c} \right]^k \right\}
\theta(k-2).
\label{eq:APkER2}
\end{equation}

\noindent
Using Eq. (\ref{eq:AP2}) and the fact that the Poisson distribution satisfies
$P(K=1) = c e^{-c}$
and
$G_0(g) = e^{-(1-g)c}$,
we find that

\begin{equation}
P(i \in {\rm AP}) = 1  -  e^{-(1-g)c}  - (1-g) c e^{-c},
\label{eq:AP2ER}
\end{equation}

\noindent
where $g$ is given by Eq. (\ref{eq:g(c)}).
Using Eq. (\ref{eq:APL1}) we obtain the probability that
a random node on the giant component of an ER network is an AP, which is
given by

\begin{equation}
P(i \in {\rm AP} | {\rm GC}) =
1 - e^{-(1-g)c}.
\label{eq:APL1ER}
\end{equation}

\noindent
Similarly, from Eq. (\ref{eq:APL0}) we obtain the probability that
a random node which resides on one of the finite components is an AP,
which is 

\begin{equation}
P(i \in {\rm AP} | {\rm FC}) =
1 - e^{-(1-g)c} - c e^{-c}.
\label{eq:APL0ER}
\end{equation}

\begin{figure}
\includegraphics[width=7cm]{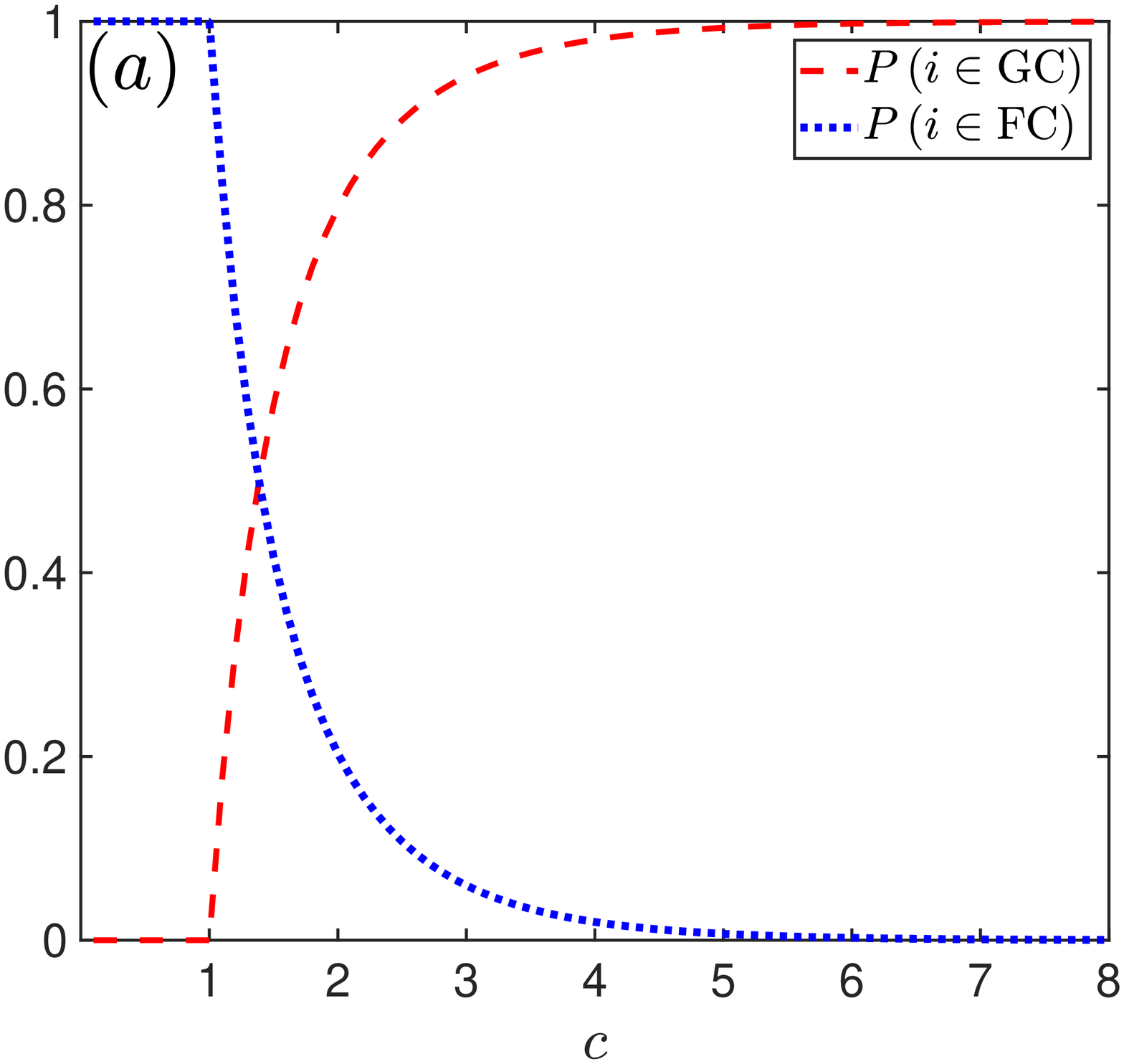} 
\includegraphics[width=7cm]{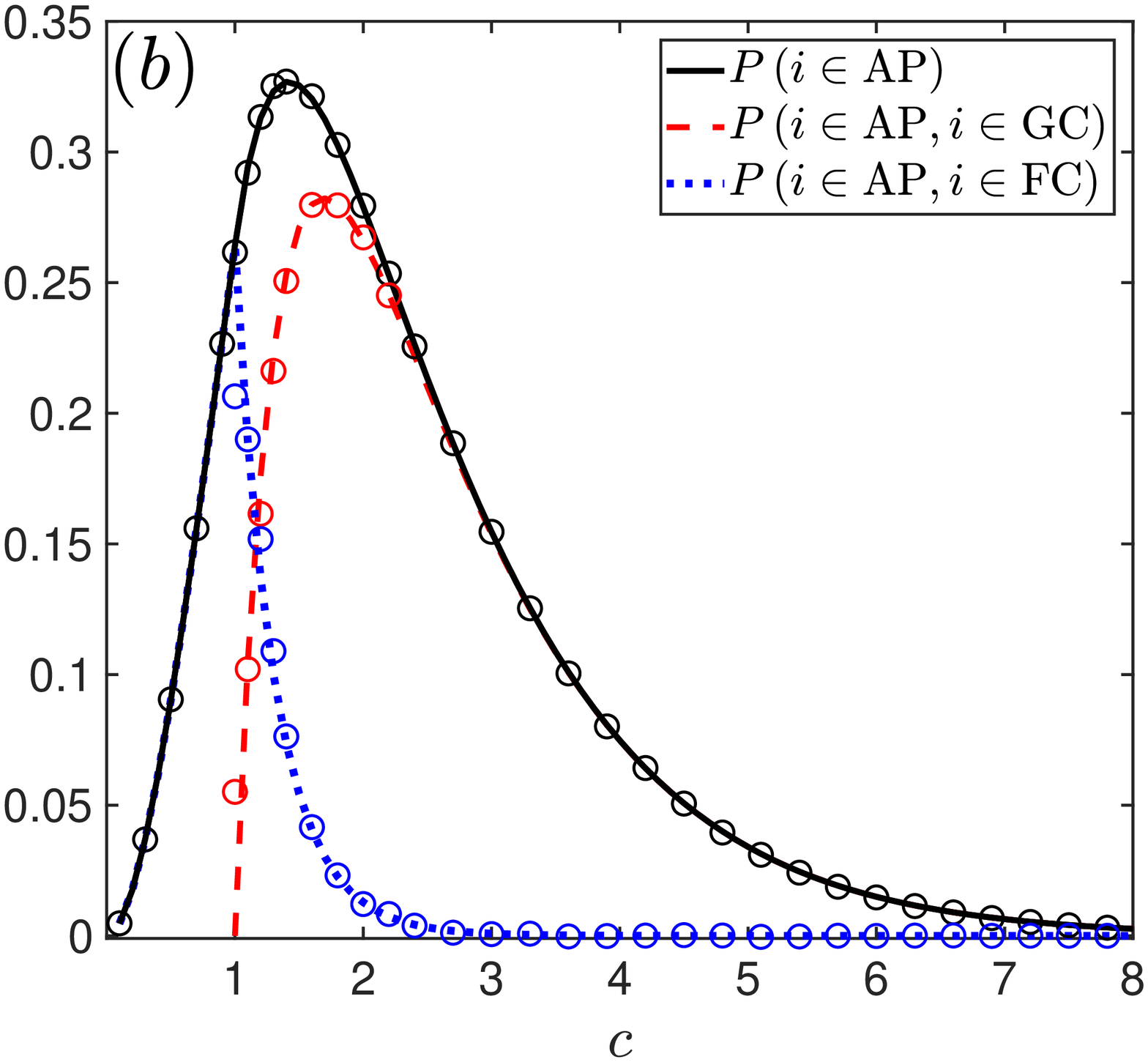} 
\caption{
(Color online)
(a) The probability $P(i \in {\rm GC}) = g$ (dashed line),
that a random node in an ER network resides on the giant component,
obtained from Eq. (\ref{eq:g(c)}),
and the probability $P(i \in {\rm FC})=1-g$ (dotted line) that such node resides on
one of the finite components;
(b) The probability
$P(i \in {\rm AP})$ 
that a random node in an ER network is an AP
(solid line),
as a function of $c$, 
obtained from Eq. (\ref{eq:AP2ER}),
the probability
$P(i \in {\rm AP} , i \in {\rm GC}) = P(i \in {\rm AP}|{\rm GC}) P(i \in {\rm GC})$,
obtained from Eq. (\ref{eq:APL1ER}),
that a randomly selected node in the network
is an AP that resides in the giant component
(dashed line)
and the probability
$P(i \in {\rm AP} , i \in {\rm FC}) = P(i \in {\rm AP}|{\rm FC}) P(i \in {\rm FC})$,
obtained from Eq. (\ref{eq:APL0ER}),
that a randomly selected node in the network is an AP that resides in one of the finite
components (dotted line).
The analytical results
are found to be in very good agreement with 
the results of
computer simulations
(circles),
performed for an ensemble of ER networks
of $N=1000$ nodes.
It is found that in the limit of sparse networks,
the probability
$P(i \in {\rm AP})$
exhibits a peak around
$c \simeq 1.5$,
where the network is most sensitive 
to fragmentation due to the deletion of APs. 
}
\label{fig:2}
\end{figure}

In Fig. \ref{fig:2}(a) we present the probability $P(i \in {\rm GC}) = g$ (dashed line),
that a random node in an ER network resides on the giant component,
obtained from Eq. (\ref{eq:g(c)}),
and the probability 
$P(i \in {\rm FC}) = 1-g$ (dotted line)
that such node resides on one of the finite components.
In Fig. \ref{fig:2}(b) we present analytical results 
for the probability
$P(i \in {\rm AP})$ (solid line) that a random node in
an ER network is an AP,
as a function of $c$.
We also present
the probability
$P(i \in {\rm AP}, i \in {\rm GC})$
that a random node is an AP which resides on the giant component 
(dashed line),
and the probability
$P(i \in {\rm AP}, i \in {\rm FC})$
that a random node is an AP which resides in one of the finite
components (dotted line).
The analytical results
are found to be in very good agreement with 
the results of
computer simulations
(circles),
performed for an ensemble of ER networks
of $N=1000$ nodes.
It is found that in the subcritical regime,
the probability
$P(i \in {\rm AP})$
increases monotonically as a function of $c$.
This is due to the fact that as $c$ is increased from $0$ to $1$, the
finite tree components become larger and the fraction
of nodes of degrees $k \ge 2$ quickly increases.
For $c > 1$ the contribution of the finite components to
$P(i \in {\rm AP})$ 
sharply decreases while the giant component
becomes dominant. 
Just above the percolation threshold, the giant component also
includes a large fraction of APs and therefore the
probability
$P(i \in {\rm AP})$
continues to increase.
It reaches its maximal value around
$c \simeq 1.5$ and then gradually decreases
as $c$ is increased further.
The decrease in 
$P(i \in {\rm AP})$
is due to the fact that as the network becomes
more dense more and more cycles are formed,
thus reducing the number of APs.
These results imply that slightly above the percolation
threshold, ER networks are most sensitive to
disintegration due to the deletion of APs.

Using Eq. (\ref{eq:L1AP}) we obtain the probability that a randomly selected
AP in an ER network resides on the giant component, which is given by

\begin{equation}
P(i \in {\rm GC} | {\rm AP}) = 
\left[ \frac{ 1-g-e^{-c} }{ 1-g-e^{-c}  -  (1-g)^2 c e^{-c}  } \right] g.
\end{equation}

\noindent
The complementary probability,
$P(i \in {\rm FC} | {\rm AP}) = 1 - P(i \in {\rm GC} | {\rm AP})$,
is given by

\begin{equation}
P(i \in {\rm FC} | {\rm AP}) 
= 
\left[ \frac{ 1-g- e^{-c} - (1-g) c e^{-c} }{  1 - g - e^{-c}  - (1-g)^2 ce^{-c}  } \right] (1-g).
\end{equation}

Using Eq. (\ref{eq:kAP}) we obtain the
degree distribution of APs in ER networks,
which is given by

\begin{equation}
P(K=k | {\rm AP}) =
\left[  \frac{ (1 - g^k) \theta(k-2) }
{1  - e^{-c(1-g)} - (1-g)ce^{-c}  }  \right]
\frac{e^{-c} c^k}{k!}.
\label{eq:kAPER}
\end{equation}

\noindent
Using Eq. (\ref{eq:kAPL1}) we obtain the
degree distribution of APs which reside on the giant component,
which is given by

\begin{equation}
P(K=k | {\rm AP},{\rm GC}) =
\left[ \frac{1 - g^k - (1-g)^k}{g - e^{-(1-g)c} + e^{-c} } \right]
\frac{e^{-c} c^k}{k!} \theta(k-2).
\label{eq:kAPL1ER}
\end{equation}

\noindent
Using Eq. (\ref{eq:kAPL0}) we obtain 
the degree distribution of APs which reside on the finite components,
which takes the form

\begin{equation}
P(K=k | {\rm AP},{\rm FC}) =
- \left[ \frac{ (1-g)^{k-1} \theta(k-2) }{1 - e^{-(1-g)c} -c e^{-c} } \right]
\frac{e^{-c} c^k}{k!}.
\label{eq:kAPL0ER}
\end{equation}

\begin{figure}
\includegraphics[width=7cm]{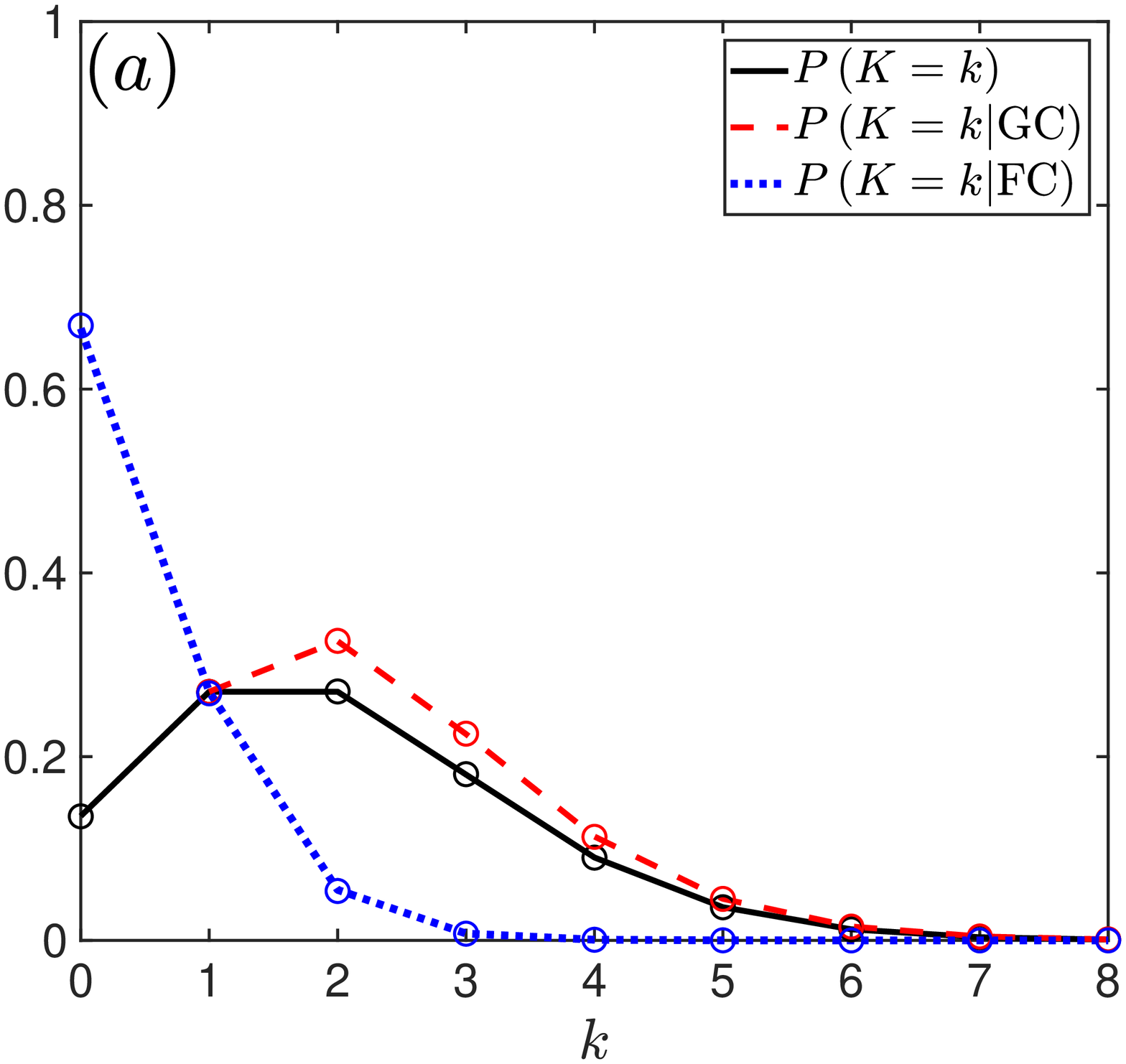}
%\\
\includegraphics[width=7cm]{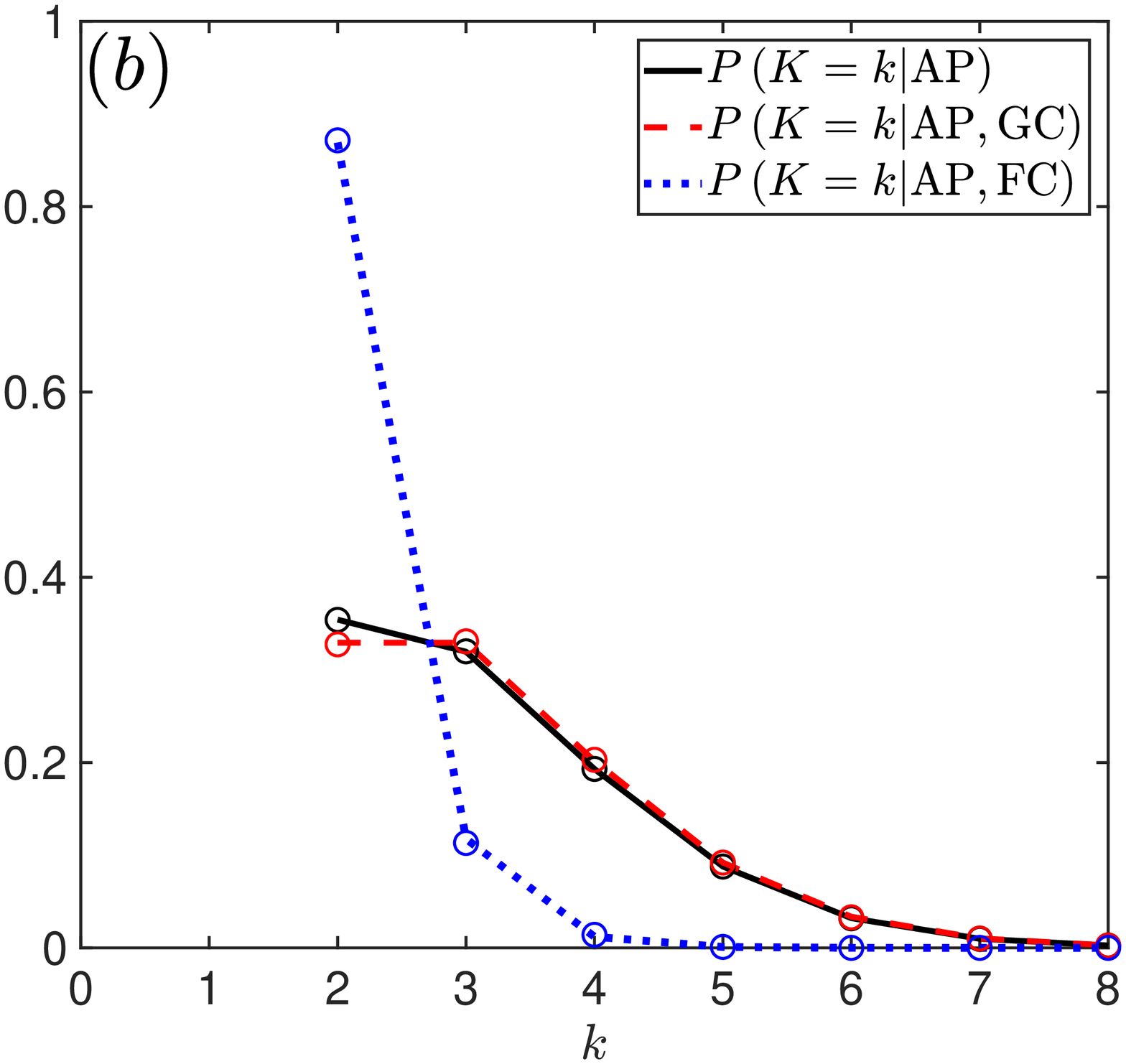}
\caption{
(Color online)
(a) Analytical results for the degree distribution
$P(K=k)$ 
of an ER network with $c=2$
(solid line) ,
the degree distribution,
$P(K=k | {\rm GC})$,
of the giant component (dashed line),
obtained from Eq. (\ref{eq:APL1ER}), 
and the degree distribution,
$P(K=k | {\rm FC})$,
of the finite components (dotted line),
obtained from Eq. (\ref{eq:APL0ER}); 
(b) Analytical results for the degree distribution
$P(K=k | {\rm AP})$ (solid line) 
of APs in an ER network with $c=2$,
obtained from Eq. (\ref{eq:kAPER})
the degree distribution
$P(K=k | {\rm AP},{\rm GC})$ 
of APs in the giant component (dashed line),
obtained from Eq. (\ref{eq:kAPL1ER})
and the degree distribution
$P(K=k | {\rm AP},{\rm FC})$ 
of APs in the finite components
(dotted line),
obtained from Eq. (\ref{eq:kAPL0ER}).
The analytical results are in very good agreement with the
results of computer simulations (circles).
}
\label{fig:3}
\end{figure}

In Fig. \ref{fig:3}(a) we present analytical results 
for the degree distribution
$P(K=k)$  of an ER network with $c=2$ (solid line),
the degree distribution
$P(K=k | {\rm GC})$ 
of the giant component (dashed line)
and the degree distribution
$P(K=k | {\rm FC})$ 
of the finite components (dotted line).
These results are found to be in very good
agreement with the results of computer simulations (circles).
In Fig. \ref{fig:3}(b) we present analytical results
for the degree distribution 
$P(K=k | {\rm AP})$ (solid line)
of APs in an
ER network with $c=2$.
We also show the degree distribution 
$P(K=k | {\rm AP},{\rm GC})$ (dashed line),
of APs which reside on the giant component
and the degree distribution 
$P(K=k | {\rm AP},{\rm FC})$ (dotted line)
of APs which reside on the finite components.
For a given value of $c$, nodes of higher degree are more likely to 
be APs.
This is due to the fact that each one of the $k$ neighbors 
of a node $i$ of degree $k$ may detach from the network component
on which $i$ resides upon deletion of node $i$.
In order for node $i$ to be an AP it is sufficient that one of its neighbors
will detach upon deletion of $i$.

Using Eq. (\ref{eq:EkAP}) we obtain
the mean degree of the APs in ER networks,
which is given by

\begin{equation}
\mathbb{E}[K|{\rm AP}] =
\left[  \frac{ 1-g e^{-c(1-g)} - (1-g) e^{-c} }
{ 1 - e^{-c(1-g)} - (1-g) c e^{-c} } \right]
c.
\label{eq:EkAPER}
\end{equation}

\begin{figure}
\includegraphics[width=7cm]{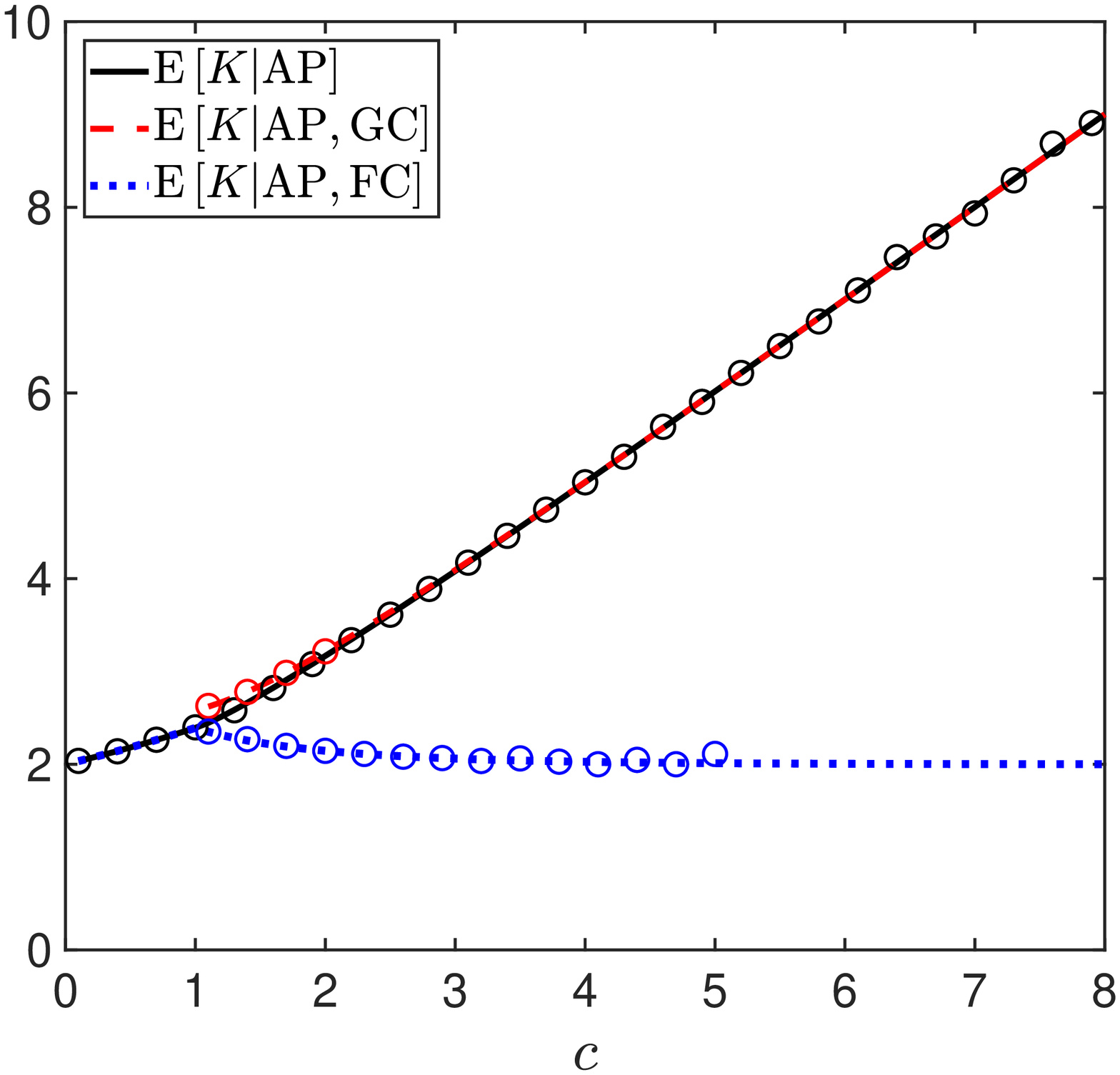}
\caption{
(Color online)
Analytical results for the mean degree, 
$\mathbb{E}[K | {\rm AP}]$,
of APs as a function of $c$ in an ER network
(solid line),
obtained from Eq. (\ref{eq:EkAPER}),
the mean degree,
$\mathbb{E}[K | {\rm AP},{\rm GC}]$,
of APs in the giant component (dashed line),
obtained from Eq. (\ref{eq:EkAPL1}),
and the mean degree,
$\mathbb{E}[K | {\rm AP},{\rm FC}]$,
of APs in the finite components (dotted line)
obtained from Eq. (\ref{eq:EkAPL0}).
The analytical results 
are in very good agreement with the
results of computer simulations (circles).
}
\label{fig:4}
\end{figure}

\noindent
In Fig. \ref{fig:4} we present analytical results for
the mean degree of the APs in ER networks,
$\mathbb{E}[K|{\rm AP}]$ (solid line),
obtained from Eq. (\ref{eq:EkAPER}),
as a function of $c$.
We also present analytical results for the mean degree of APs which reside on the
giant component,
$\mathbb{E}[K|{\rm AP},{\rm GC}]$ (dashed line)
and the mean degree of APs which reside on the
finite components,
$\mathbb{E}[K|{\rm AP},{\rm FC}]$ (dotted line).
These results are obtained from Eqs. 
(\ref{eq:EkAPL1}) and (\ref{eq:EkAPL0}),
respectively, by inserting the expressions for
$g$, $G_1(x)$ and $P(K=1)$ for ER networks.
The analytical results are in very good agreement with
the corresponding results obtained from computer
simulations (circles).

The distribution of ranks of random nodes that reside on the
giant component of an ER network can be obtained from
Eq. (\ref{eq:rL1}), by inserting the Poisson distribution for
$P(K=k)$ and using the condition $g=\tilde g$. Performing
the summation, we obtain

\begin{equation}
P(R=r|{\rm GC}) = \frac{e^{-(1-g)c}  [(1-g)c]^r}{r!}.
\label{eq:rL1ER} 
\end{equation}

\noindent
Thus, the rank distribution on the giant component is a Poisson
distribution, whose mean is given by

\begin{equation}
\mathbb{E}[R|{\rm GC}] = (1-g) c.
\label{eq:ErL1ER}
\end{equation}

\noindent
Using Eq. (\ref{eq:rL0}) we obtain the
distribution of articulation ranks of random nodes which reside on the
finite components, which is given by

\begin{equation}
P(R=r | {\rm FC}) =
\left\{
\begin{array}{ll}
e^{-(1-g)c} +  c e^{-c}   & \ \  r=0 \\
(1 -  g)^{r}   \frac{e^{-c} c^{r+1}}{(r+1)!}   & \ \   r \ge 1.
\end{array}
\right.
\label{eq:rL0ER}
\end{equation}

\noindent
The overall distribution of articulation ranks is given by

\begin{equation}
P(R=r) =
\left\{
\begin{array}{ll}
e^{-(1-g)c} + (1 - g) c e^{-c}   & \ \  r=0 \\
e^{-(1-g)c} 
\left\{  g \frac{[(1-g)c]^r}{r!} 
+ (1-g) \frac{[(1-g)c]^{r+1}}{(r+1)!} \right\}  & \ \   r \ge 1.
\end{array}
\right.
\label{eq:rER}
\end{equation}

The mean articulation rank of nodes which reside on the finite components is

\begin{equation}
\mathbb{E}[R| {\rm FC}] =
e^{-(1-g)c} + c e^{-gc} - 1.
\label{eq:ErL0ER}
\end{equation}

\noindent
The mean rank of the whole network is given by

\begin{equation}
\langle R \rangle =
\mathbb{E}[R|{\rm GC}] P(i \in {\rm GC})
+
\mathbb{E}[R|{\rm FC}] P(i \in {\rm FC}).
\label{eq:RER}
\end{equation}

\noindent
Inserting 
$\mathbb{E}[R|{\rm GC}]$
and
$\mathbb{E}[R|{\rm FC}]$
from Eqs. 
(\ref{eq:ErL1ER})
and
(\ref{eq:ErL0ER}),
respectively into Eq. (\ref{eq:RER})
we obtain

\begin{equation}
\langle R \rangle = e^{-c} + (c-1) (1-g).
\label{eq:ErER}
\end{equation}

\begin{figure}
\includegraphics[width=7cm]{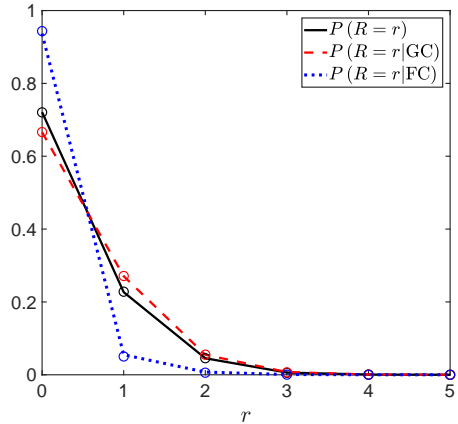}
\caption{
(Color online)
Analytical results for the distribution 
$P(R=r)$ 
of the articulation ranks of nodes in an ER network
with $c=2$ (solid line),
obtained from Eq. (\ref{eq:rER})
the distribution
$P(R=r | {\rm GC})$ 
of the articulation ranks of nodes in the giant component
(dashed line),
obtained from Eq. (\ref{eq:rL1ER})
and the distribution
$P(R=r | {\rm FC})$ 
of the articulation ranks of nodes in the finite components
(dotted line),
obtained from Eq. (\ref{eq:rL0ER}).
The analytical results
are in very good agreement with the
results of computer simulations (circles).
}
\label{fig:5}
\end{figure}

In Fig. \ref{fig:5} we present analytical results for
the rank distribution $P(R=r)$
(solid line) of an ER network with $c=2$,
the rank distribution $P(R=r | {\rm GC})$ (dashed line)
of the giant component and the rank distribution
$P(R=r | {\rm FC})$ (dotted line)
of the finite components.
We also present the corresponding results obtained
from computer simulations (circles), which are in
very good agreement with the analytical results.

\begin{figure}
\includegraphics[width=7cm]{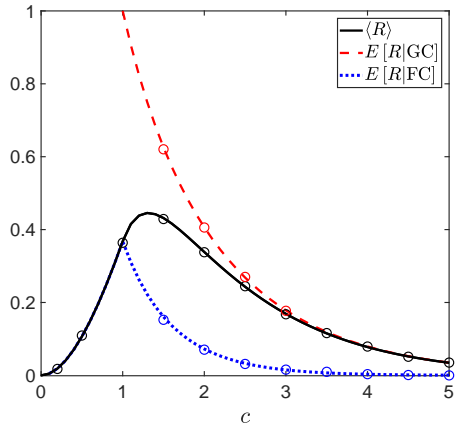}
\caption{
(Color online)
Analytical results for the mean articulation rank,
$\langle R \rangle$,  
of an ER network,
as a function of the mean degree, $c$
(solid line),
obtained from Eq. (\ref{eq:ErER}).
The mean articulation rank
$\mathbb{E}[R|{\rm GC}]$ 
of the nodes that reside in the giant component
(dashed line),
obtained from Eq. (\ref{eq:ErL1ER}),
and the mean articulation rank
$\mathbb{E}[R|{\rm FC}]$ 
of the nodes that reside in the finite components
(dotted line),
obtained from Eq. (\ref{eq:ErL0ER}),
are also shown.
The analytical results 
are in very good agreement with the
results of computer simulations (circles).
}
\label{fig:6}
\end{figure}

In Fig. \ref{fig:6} we present
analytical results for the mean articulation rank,
$\langle R \rangle$,  
of an ER network (solid line),
as a function of the mean degree, $c$,
the mean articulation rank
$\mathbb{E}[R|{\rm GC}]$ 
of the nodes which reside in the giant component
(dashed line)
and the mean articulation rank
$\mathbb{E}[R|{\rm FC}]$ 
of the nodes which reside in the finite components
(dotted line).
The analytical results 
are in very good agreement with the
results of computer simulations (circles).
At the percolation threshold ($c=1$), the mean articulation
rank of the nodes which reside on the giant component is 
$\mathbb{E}[R|{\rm GC}]=1$,
and it decreases monotonically as $c$ is increased.
The mean articulation rank of the nodes which reside
on the finite components exhibits a cusp at the percolation threshold,
and it sharply decreases on both sides.
The mean articulation rank of the whole network, which is presented only
for the supercritical regime in which the giant and the finite components
coexist, increases above the percolation threshold and then decreases as
$c$ is increased.

Inserting the properties of ER networks in Eq. (\ref{eq:Stub1}),
we find that the probability that a random $k=2$ AP that resides on the GC
is a tube is

\begin{equation}
P({\rm Tube}) = 1 + \frac{c e^{-c}}{(1-c e^{-c})W(-ce^{-c})}.
\label{eq:TubeER}
\end{equation}

\begin{figure}
\includegraphics[width=7cm]{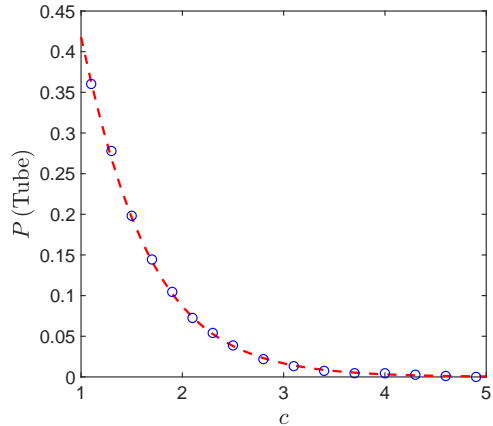}
\caption{
(Color online)
Analytical results (dashed line) for the
probability $P({\rm Tube})$ that a random $k=2$ AP on the GC 
of an ER network is a tube,
as a function of the mean degree $c$, 
obtained from Eq. (\ref{eq:TubeER}).
The analytical results are in excellent agreement with the
results obtained from computer simulations (circles).
}
\label{fig:7}
\end{figure}

\noindent
In Fig. \ref{fig:7} we present the probability $P({\rm Tube})$ as a function of the
mean degree, $c$, for ER networks. It is found that $P({\rm Tube})$ is a monotonically
decreasing function of $c$. This implies that as the network becomes more dense
the trees that detach upon deletion of APs become of simpler topologies.

\subsection{Configuration model networks with exponential degree distributions}

Consider a configuration model network with an exponential degree
distribution of the form $P(K=k) \sim e^{- \alpha k}$,
where $k_{\rm min} \le k \le k_{\rm max}$.
In case that $k_{\rm min} \ge 2$, it can be shown that
$g = \tilde g =1$ and there are no APs.
Here we consider the case of
$k_{\rm min}=1$ 
and 
$k_{\rm max} = \infty$.
In this case it is convenient to express the degree distribution 
in the form

\begin{equation}
P(K=k) = \frac{1}{c-1} \left( \frac{c-1}{c} \right)^k,
\label{eq:exp}
\end{equation}

\noindent
where
$c= \langle K \rangle$ 
is the mean degree.
In order to find the properties of APs in such networks,
we first calculate the parameters $\tilde g$ and $g$. 
Inserting the exponential degree distribution of 
Eq. (\ref{eq:exp}) 
into the generating function
$G_1(x)$, given by Eq. (\ref{eq:G1}),
we obtain

\begin{equation}
G_1(x) = \frac{1}{ [ c - (c-1) x ]^2 }.
\label{eq:G1exp}
\end{equation}

\noindent
Inserting $G_1(x)$ from Eq. (\ref{eq:G1exp}) into 
Eq. (\ref{eq:tg}) and
solving for $\tilde g$, 
we find that
for $c > 3/2$ there is a non-trivial solution of the form

\begin{equation}
\tilde g = \frac{c-3}{2(c-1)} + \frac{1}{2} \sqrt{ \frac{c+3}{c-1} }.
\label{eq:tgXP}
\end{equation}

\noindent
Inserting this result into Eq. (\ref{eq:APk}), 
we obtain

\begin{equation}
P(i \in {\rm AP} | k) = 
\left\{ 
1 - \left[ \frac{c-3}{2(c-1)} + \frac{1}{2} \sqrt{ \frac{c+3}{c-1} } \right]^k   
\right\} \theta(k-2).
\end{equation}

\noindent
Inserting the exponential degree distribution of Eq. (\ref{eq:exp}) into
Eq. (\ref{eq:G0}), we obtain

\begin{equation}
G_0(x) = \frac{x}{ c - (c-1)x }.
\end{equation}

\noindent
Inserting the expression for $\tilde g$ 
from Eq. (\ref{eq:tgXP})
into $G_0(1-\tilde g)$ in
Eq. (\ref{eq:G0}),
we find that for $c > 3/2$

\begin{equation}
g = 
\frac{ 3c }{ 2 (c-1) }
-\frac{c}{2(c-1)} \sqrt{ \frac{c+3}{c-1} }.
\label{eq:gXP}
\end{equation}

\noindent
Thus, it is found that the configuration model network with an exponential
degree distribution exhibits a percolation transition at $c_0 = 3/2$,
below which $g=\tilde g =0$ and above which
$\tilde g > 0$ is given by Eq. (\ref{eq:tgXP}) and
$g > 0$ is given by Eq. (\ref{eq:gXP}).
Inserting this result into Eq. (\ref{eq:AP2}),
we obtain the probability that a random node is an AP, 
which is given by

\begin{equation}
P(i \in {\rm AP}) =
1 - \frac{\tilde g}{c-(c-1)\tilde g}
- \frac{1-\tilde g}{c}.
\end{equation}

\noindent
Using Eq. (\ref{eq:APL1}), we obtain the probability that
a random node on the giant component is an AP, 
which is given by

\begin{equation}
P(i \in {\rm AP} | {\rm GC}) =
1 - \frac{\tilde g}{ [ c - (c-1) \tilde g ] g }.
\end{equation}

\noindent
Similarly, using Eq. (\ref{eq:APL0}), we obtain the probability 
that a random node which resides on one of the finite components
is an AP, which is

\begin{equation}
P(i \in {\rm AP} | {\rm FC}) =
1 - \frac{1 - \tilde g}{ (1-g) c},
\end{equation}

\noindent
where $\tilde g$ is given by Eq. (\ref{eq:tgXP})
and $g$ is given by Eq. (\ref{eq:gXP}).

The probability that a randomly selected AP resides on the giant component
is given by

\begin{equation}
P(i \in {\rm GC} | {\rm AP}) = 
\frac{ c^2 g (1 - \tilde g) - c \tilde g (1-g) }
{(1-\tilde g)[c^2 - c(1-\tilde g) - \tilde g]},
\end{equation}

\noindent
while the probability that such node resides on one
of the finite components is 
$P(i \in {\rm FC} | {\rm AP}) =
1 - P(i \in {\rm GC} | {\rm AP})$.
Using Eq. (\ref{eq:kAP}) we obtain the degree distribution
of the APs, which is given by

\begin{equation}
P(K=k | {\rm AP}) = 
\left( \frac{ 1-\tilde g^k }{1 - \tilde g} \right)
\left[ \frac{c + \tilde g -c \tilde g }{(c + \tilde g)(c-1) } \right]
\left( \frac{c-1}{c} \right)^{k-1} \theta(k-2).
\end{equation}

\noindent
The degree distribution of APs which reside on the giant component
is given by

\begin{equation}
P(K=k| {\rm AP},{\rm GC}) =
\frac{c}{(c-1)g}
\left[  
\frac{(c+\tilde g - c \tilde g) (1-\tilde g^k)  }{ (1-\tilde g)(c+\tilde g)(c-1) } 
\right]
\left( \frac{c-1}{c} \right)^k \theta(k-2),
\end{equation}

\noindent
while the degree distribution of APs which reside on the
finite components is

\begin{equation}
P(K=k| {\rm AP},{\rm FC}) =
\frac{c}{(c-1)}
\left[  \frac{ (1-\tilde g)^k \theta(k-2) }
{(1 - g)c - (1-\tilde g)} \right]
\left( \frac{c-1}{c} \right)^k.
\end{equation}

\noindent
From Eq. (\ref{eq:EkAP}) we obtain the mean degree of the APs,
which is given by

\begin{equation}
\mathbb{E}[K | {\rm AP}] =
1 + \frac{c(c-1)}{c+\tilde g} 
+ \frac{c}{c(1-\tilde g) + \tilde g}.
\end{equation}

Carrying out the summation in Eq. (\ref{eq:rL1}) with $P(K=k)$
given by Eq. (\ref{eq:exp}), we obtain

\begin{equation}
P(R=r | {\rm GC}) =
\frac{1}{(c-1)g} \left[ \frac{ (1-\tilde g)  (c-1)}{c}  \right]^r
\left[ \left( \frac{c}{  c - \tilde g c + \tilde g} \right)^{r+1} 
-1 \right].
\end{equation}

Inserting the properties of configuration model networks with exponential
degree distributions  in Eq. (\ref{eq:Stub1}),
we find that the probability that a random $k=2$ AP that resides on the GC
is a tube is

\begin{equation}
P({\rm Tube}) = 1 - 
\frac{c}{[c^3-2(c-1)] \left[ 1 -  \frac{c-3}{2(c-1)} - \frac{1}{2} \sqrt{\frac{c+3}{c-1}}
 \right]}.
\end{equation}

\noindent
It is found that $P({\rm Tube})$ is a monotonically
decreasing function of $c$. This implies that as the network becomes more dense
the trees that detach upon deletion of APs become of simpler topologies.

\subsection{Configuration model networks with power-law degree distributions}

Consider a configuration model network with a power-law degree distribution
of the form 
$P(K=k) \sim k^{-\gamma}$,
where 
$k_{\rm min} \le k \le k_{\rm max}$.
Here we focus on the case of $\gamma > 2$, in which the
mean degree, 
$\langle K \rangle$, 
is bounded even for 
$k_{\rm max} \rightarrow \infty$.
Power-law distributions do not exhibit a typical scale, and are therefore
referred to as scale-free networks.
The normalized degree distribution is given by

\begin{equation}
P(K=k) = \frac{ k^{-\gamma} }{ \zeta(\gamma,k_{\rm min}) - \zeta(\gamma,k_{\rm max}+1) },
\label{eq:PLnorm}
\end{equation}

\noindent
where $\zeta(\gamma,k)$ is the Hurwitz zeta function 
\cite{Olver2010}.
For $\gamma \le 2$ the mean degree diverges when 
$k_{\rm max} \rightarrow \infty$.
For $2 < \gamma \le 3$ the mean degree is bounded while the second moment,
$\langle K^2 \rangle$, diverges in the limit of
$k_{\rm max} \rightarrow \infty$.
For $\gamma > 3$ both moments are bounded.
For $\gamma > 2$ and
$k_{\rm min} \ge 2$ 
(where nodes of degrees $0$ and $1$ do not exist),
one can show that
$\langle K^2 \rangle > 2 \langle K \rangle$
namely the Molloy and Reed criterion is satisfied
and the network exhibits a giant component
\cite{Molloy1995}. 
Moreover, in this case
the giant component encompasses the entire network,
namely $g = \tilde g = 1$
\cite{Bonneau2017}.
In this case there are no APs in the network and
$P(i \in {\rm AP})=0$.
Here we focus on the case of $k_{\rm min}=1$,
in which there are APs in the network.
In this case, for $\gamma > 2$
the mean degree is given by

\begin{equation}
\langle K \rangle = 
\frac{ \zeta(\gamma-1) - \zeta(\gamma-1,k_{\rm max}+1) }
{ \zeta(\gamma) - \zeta(\gamma,k_{\rm max}+1) },
\label{eq:Kmsf}
\end{equation}

\noindent
where $\zeta(\gamma) = \zeta(\gamma,1)$ is the Riemann zeta function.
To obtain $\langle K \rangle$ for $\gamma=2$ we take the limit 
$\gamma \rightarrow 2^{+}$ in Eq. (\ref{eq:Kmsf}).
We obtain

\begin{equation}
\langle K \rangle_{\gamma=2} = 
\frac{ H_{k_{\rm max}}}{ H_{k_{\rm max}}^{(2)} },
\label{eq:cmax}
\end{equation}

\noindent
where $H_{n}$ is the $n$th harmonic number
and $H_n^{(m)}$ is the generalized $n$th harmonic number
of order $m$
\cite{Srivastava2012}.
For $k_{\rm max} \gg 1$, it satisfies

\begin{equation}
\langle K \rangle_{\gamma=2} \simeq \frac{6}{\pi^2} \ln k_{\rm max}.
\end{equation}

\noindent
The mean degree $\langle K \rangle$ is a monotonically decreasing function
of the exponent $\gamma$. Therefore, $\langle K \rangle_{\gamma=2}$ is
the largest possible value of $\langle K \rangle$ that can be obtained for
a given value of $k_{\rm max}$.

The second moment of the degree distribution is

\begin{equation}
\langle K^2 \rangle = 
\frac{ \zeta(\gamma-2) - \zeta(\gamma-2,k_{\rm max}+1) }
{ \zeta(\gamma) - \zeta(\gamma,k_{\rm max}+1) }.
\label{eq:K2msf}
\end{equation}

\noindent
Inserting the degree distribution of Eq. (\ref{eq:PLnorm})
with $k_{\rm min}=1$
into Eqs. 
(\ref{eq:G0})
and
(\ref{eq:G1})
we obtain

\begin{equation}
G_0(x) = \frac{ {\rm Li}_{\gamma}(x) - x^{k_{\rm max}+1}  \Phi(x,\gamma,k_{\rm max}+1)}
{ \zeta(\gamma) - \zeta(\gamma,k_{\rm max}+1) },
\end{equation}

\noindent
and

\begin{equation}
G_1(x) = \frac{ {\rm Li}_{\gamma-1}(x) - x^{k_{\rm max}+1} \Phi(x,\gamma-1,k_{\rm max}+1) }
{ x [\zeta(\gamma-1) - \zeta(\gamma-1,k_{\rm max}+1)] },
\end{equation}

\noindent
where 
$\Phi(x,\gamma,k)$ is the Lerch transcendent and
${\rm Li}_{\gamma}(x)$ is the polylogarithm function
\cite{Gradshteyn2000}.
The values of the parameters $\tilde g$ and $g$ are determined by
Eqs. (\ref{eq:tg}) and (\ref{eq:g}).
Unlike the ER network and the configuration model network
with an exponential degree distribution, here we do not have
closed form analytical expressions for $g$ and $\tilde g$.
However, using the expressions above
for $G_0(x)$ and $G_1(x)$, 
the values of $g$ and $\tilde g$ can be easily obtained
from a numerical solution of 
Eqs. (\ref{eq:tg}) and (\ref{eq:g}).
To find the percolation threshold we use the
Molloy-Reed criterion, which states that at the
transition
$\langle K^2 \rangle - 2 \langle K \rangle = 0$
\cite{Molloy1995,Molloy1998}.
Inserting the expressions for 
$\langle K \rangle$ 
and
$\langle K^2 \rangle$ 
from Eqs.
(\ref{eq:Kmsf}) and (\ref{eq:K2msf}),
for $k_{\rm max}=100$
we find that the percolation threshold takes place at
$\gamma_0 \simeq 3.37876$,
where the mean degree is
$c_0 \simeq 1.21946$.

\begin{figure}
\begin{center}
\includegraphics[width=7cm]{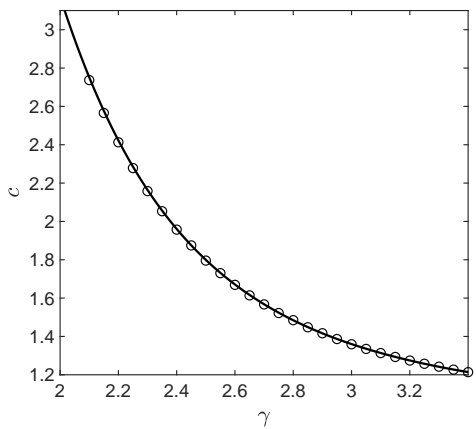} 
\end{center}
\caption{
The mean degree, 
$c= \langle K \rangle$,
as a function of the exponent $\gamma$,
for a configuration model network 
with a power-law degree
distribution, where $k_{\rm min}=1$ and $k_{\rm max}=100$.
As $\gamma$ is increased, the tail of the degree distribution 
decays more quickly and the mean degree decreases.
The analytical results (solid line),
obtained from Eq. (\ref{eq:Kmsf}),
are found to be in excellent agreement with the results of computer 
simulations (circles) performed for networks
of $N=4 \times 10^4$ nodes.
}
\label{fig:8}
\end{figure}

In Fig. \ref{fig:8} we present
the mean degree, 
$c= \langle K \rangle$,
as a function of the exponent $\gamma$,
for a configuration model network 
with a power-law degree
distribution, where $k_{\rm min}=1$ and $k_{\rm max}=100$.
As $\gamma$ is increased, the tail of the degree distribution 
decays more quickly and as a results the mean degree decreases.

\begin{figure}
\begin{center}
\includegraphics[width=6.0cm]{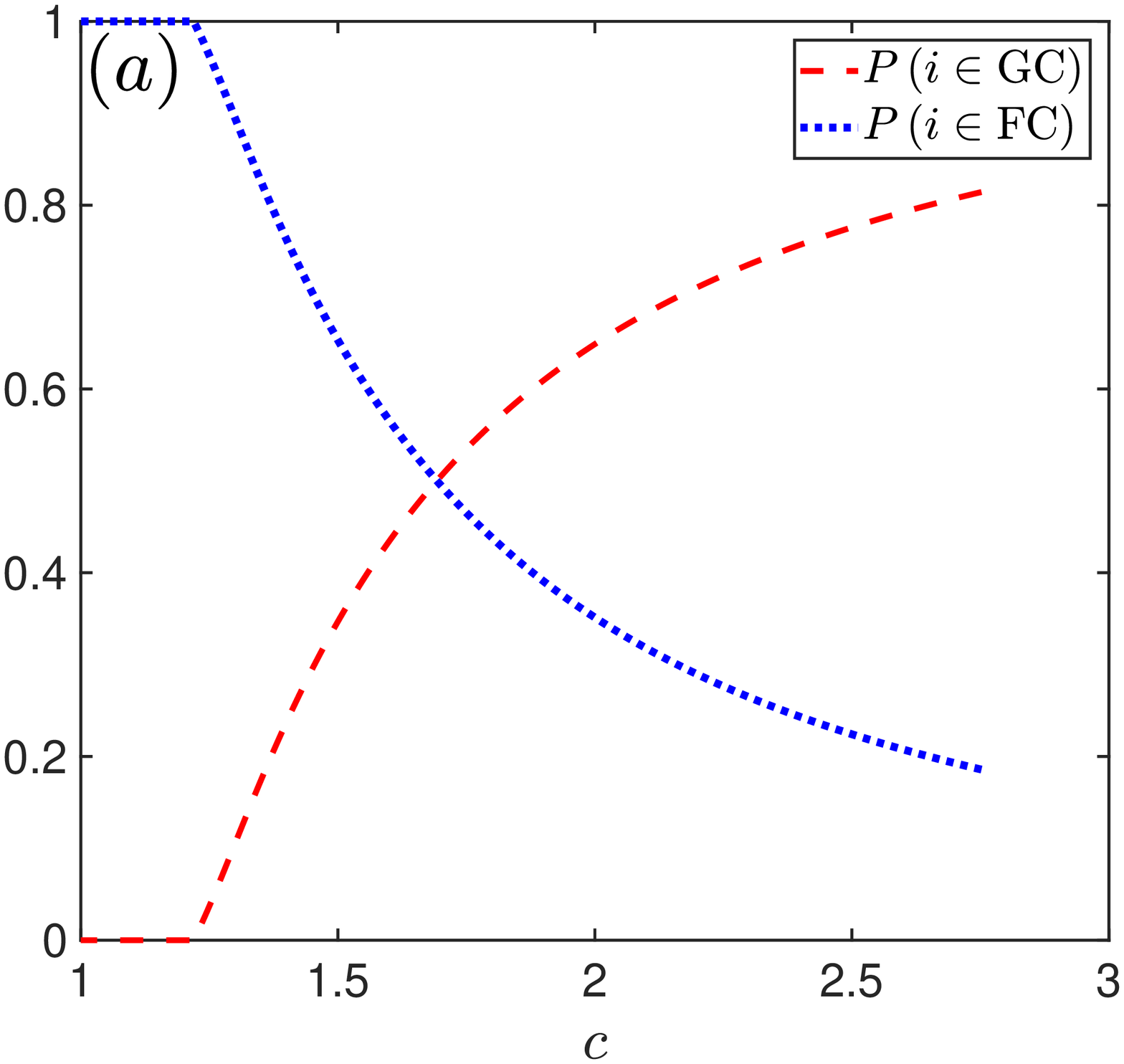} 
\includegraphics[width=6.0cm]{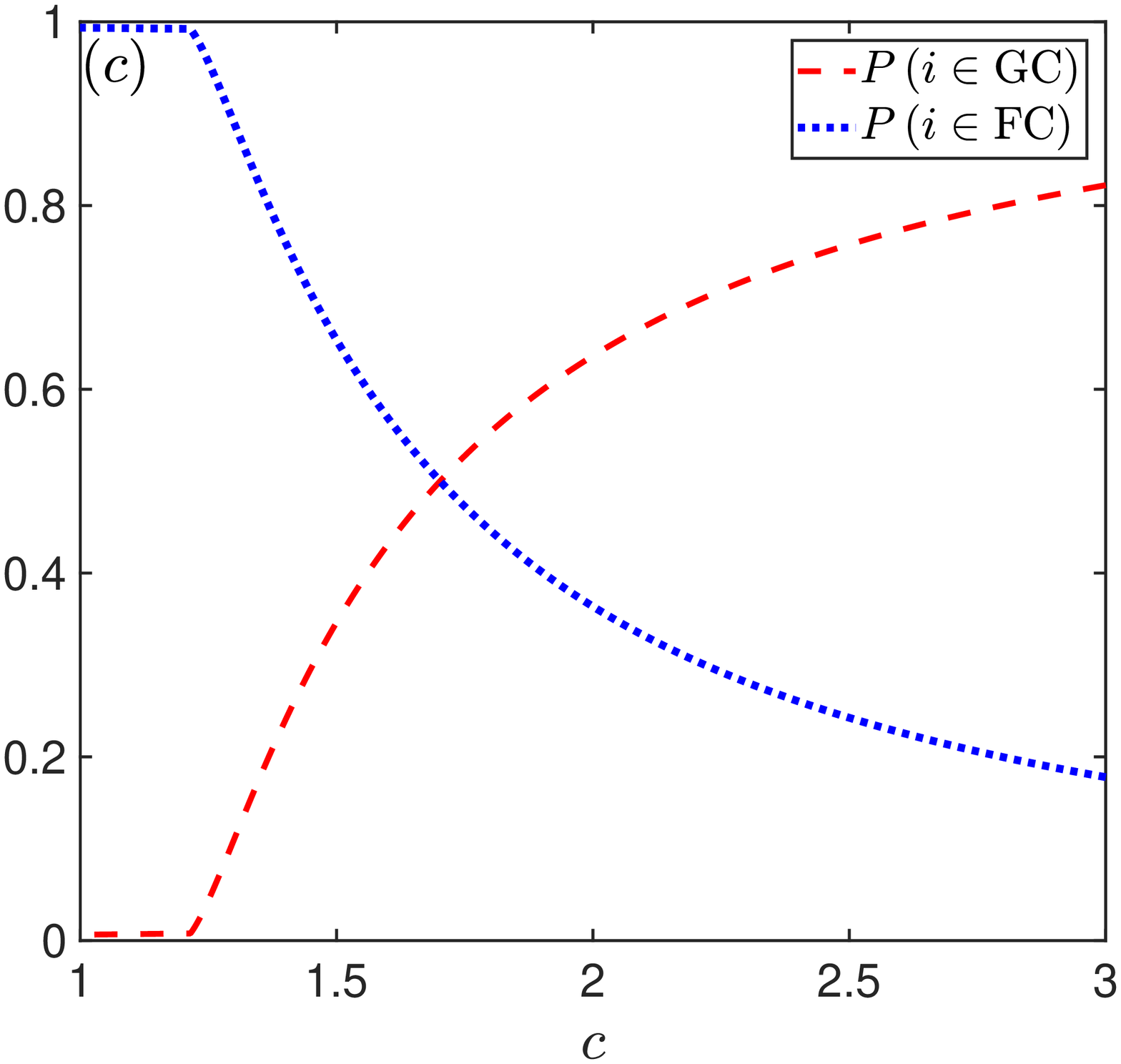} 
\\
\includegraphics[width=6.0cm]{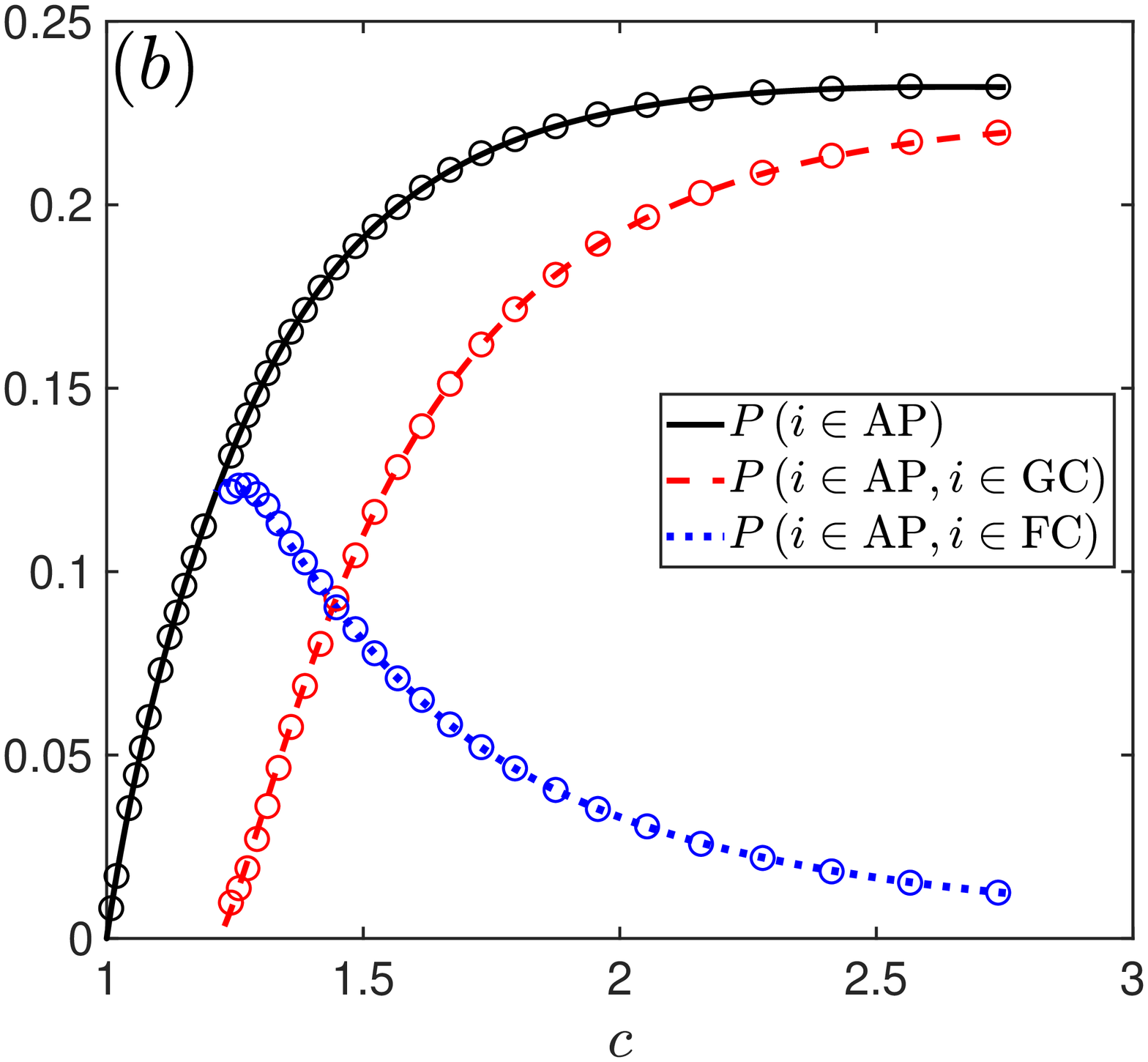}
\includegraphics[width=6.0cm]{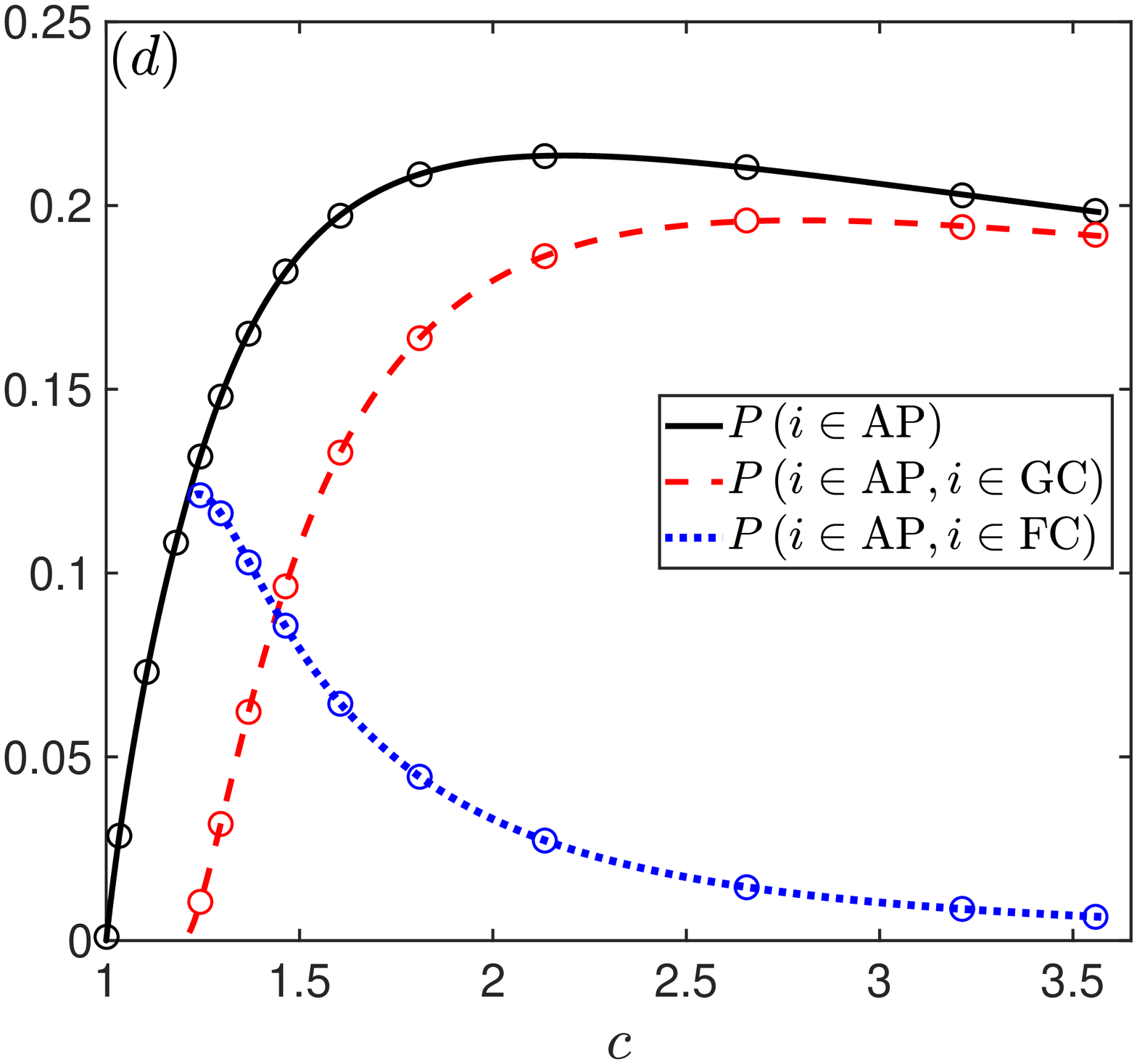} 
\end{center}
\caption{
(Color online)
(a) Analytical results for the probability,
$P(i \in {\rm GC})=g$, 
that a random node in a 
configuration model network 
with a power-law degree
distribution resides on the giant component (dashed line), 
as a function of the mean degree, $c$,
where $k_{\rm min}=1$ and $k_{\rm max}=100$.
The complementary probability,
$P(i \in {\rm FC})=1-g$,
that a randomly selected node resides on one of the finite components,
is also shown (dotted line);
(b) Analytical results for the probability, $P(i \in {\rm AP})$, 
that a randomly selected node, $i$, is an AP
(solid line)
in a configuration model network
with a power-law degree distribution,
as a function of the mean degree $c$,
obtained from Eq. (\ref{eq:iAPSF}).
The probability
$P(i \in {\rm AP}, i \in {\rm GC}) = P(i \in {\rm AP}|{\rm GC}) P(i \in {\rm GC})$,
that a randomly selected node in the network is an AP that resides in the giant
component (dashed line),
obtained from Eq. (\ref{eq:iAPL1SF}),
and the probability
$P(i \in {\rm AP}, i \in {\rm FC}) = P(i \in {\rm AP}|{\rm FC}) P(i \in {\rm FC})$,
that a randomly selected node in the network is an AP that resides in one of the
finite components (dotted line),
obtained from Eq. (\ref{eq:iAPL0SF}),
are also shown.
The analytical results are in very good agreement with the
results obtained from computer simulations (circles),
performed for networks 
of size
$N=4 \times 10^4$;
In (c) and (d) we repeat the results shown in (a) and (b), respectively,
for a larger network of size $N=10^6$ and $k_{\rm max}=1000$.
}
\label{fig:9}
\end{figure}

In Fig. \ref{fig:9}(a) we present the probability
$P(i \in {\rm GC})=g$  (dashed line)
that a random node in a configuration model network with a power-law degree
distribution where $k_{\rm min}=1$ and $k_{\rm max}=100$ 
resides on the giant component
and the probability $P(i \in {\rm FC}) = 1-g$ (dotted line)
that such node resides on one of the finite components,
as a function of the mean degree, $c$.
Using Eq. (\ref{eq:AP2}) we obtain that the
probability that a random node is an AP is given by

\begin{equation}
P(i \in {\rm AP}) =
1 - 
\frac{ (1 -\tilde g) + 
\left[ {\rm Li}_{\gamma}(\tilde g) - \tilde g^{k_{\rm max}+1}
\Phi(\tilde g,\gamma,k_{\rm max}+1) \right] }
{ \zeta(\gamma) - \zeta(\gamma,k_{\rm max}+1) }.
\label{eq:iAPSF}
\end{equation}

\noindent
Using Eq. (\ref{eq:APL1}) we express the
probability that a random node in the giant component is an AP
in the form

\begin{equation}
P(i \in {\rm AP} | {\rm GC}) =
1 - 
\frac{ 
{\rm Li}_{\gamma}(\tilde g) - \tilde g^{k_{\rm max}+1}
\Phi(\tilde g,\gamma,k_{\rm max}+1)  }
{ g [\zeta(\gamma) - \zeta(\gamma,k_{\rm max}+1)] }.
\label{eq:iAPL1SF}
\end{equation}

\noindent
Using Eq. (\ref{eq:APL0}) we obtain the probability that a random node in
one of the finite components is an AP,
which is given by

\begin{equation}
P(i \in {\rm AP} | {\rm FC}) = 
1 - \frac{1-\tilde g}{(1-g)[\zeta(\gamma)-\zeta(\gamma,k_{\rm max}+1)]}.
\label{eq:iAPL0SF}
\end{equation}

\noindent
In Fig. \ref{fig:9}(b) we present the probability 
$P(i \in {\rm AP})$ (solid line)
that a random node in a configuration model network with a power-law degree
distribution where $k_{\rm min}=1$ and $k_{\rm max}=100$ is an AP,
as a function of $c$.
We also present 
the probability 
$P(i \in {\rm AP} ,i \in {\rm GC})$ (dashed line)
that a random node is both in the giant component and is an AP
and the probability
$P(i \in {\rm AP} ,i \in {\rm FC})$ (dotted line)
that a random node resides in one of the finite
components and is an AP.
The analytical results
are found to be in very good agreement with 
the results of
computer simulations
(circles),
performed for an ensemble of configuration model networks
with a power-law degree distribution, which consist
of $N=4 \times 10^4$ nodes.
Note that the range of values of $c$ is bounded from above by
$\langle K \rangle_{\gamma=2}$, which is determined by 
$k_{\rm max}$ according to Eq. (\ref{eq:cmax}).
In Figs. \ref{fig:9}(c) and \ref{fig:9}(d) we repeat the
results of Figs. \ref{fig:9}(a) and \ref{fig:9}(b), respectively,
for a larger network with $N=10^6$ and $k_{\max}=10^3$.
Here too the agreement between the analytical results and the simulation
results (circles) is very good.
Note that the range of values of $c$ is slightly larger due to the
larger value of $k_{\rm max}$.
Interestingly, $P(i \in {\rm AP})$
exhibits a different qualitative behavior as a function of $c$
in Figs. \ref{fig:9}(b) and \ref{fig:9}(d),
namely it is monotonically increasing for $k_{\rm max}=100$
and has a local maximum for $k_{\rm max}=1000$.

The behavior of APs on the finite components, expressed by
$P(i \in {\rm AP} | {\rm FC})$
turns out to be qualitatively similar to the corresponding results
for ER networks, presented in Fig. \ref{fig:2}(b).
However, the behavior of APs on the giant component,
expressed by
$P(i \in {\rm AP} | {\rm GC})$
is markedly different.
In ER networks the fraction of APs on the
giant component reaches a maximum and then starts
to decrease as the network becomes more dense.
In contrast, in configuration model networks with a power-law
degree distribution,
$P(i \in {\rm AP} | {\rm GC})$
increases monotonically as a function of $c$.
This is due to the fact that in a power-law degree distribution
there are still many nodes of degree $k=1$ even when the mean degree, $c$,
is very large. These leaf nodes cannot reside on cycles.
As a result, other nodes which reside along the paths
leading to the leaf nodes become APs.

The degree distribution of the APs,
obtained from Eq. (\ref{eq:kAP}),
is given by

\begin{eqnarray}
P(K=k | {\rm AP}) &=&
\left\{ 1 
- 
\frac{
(1-\tilde g) 
+ 
\left[ {\rm Li}_{\gamma}(\tilde g)
- \tilde g^{k_{\rm max}+1} \Phi(\tilde g,\gamma,k_{\rm max}+1) \right] 
}
{\zeta(\gamma) - \zeta(\gamma,k_{\rm max}+1) } \right\}^{-1}
\nonumber \\
&\times&
\frac{ (1-\tilde g^k) k^{-\gamma} }
{ \zeta(\gamma) - \zeta(\gamma,k_{\rm max}+1) } \theta(k-2).
\label{kAPsf}
\end{eqnarray}

\begin{figure}
\includegraphics[width=7cm]{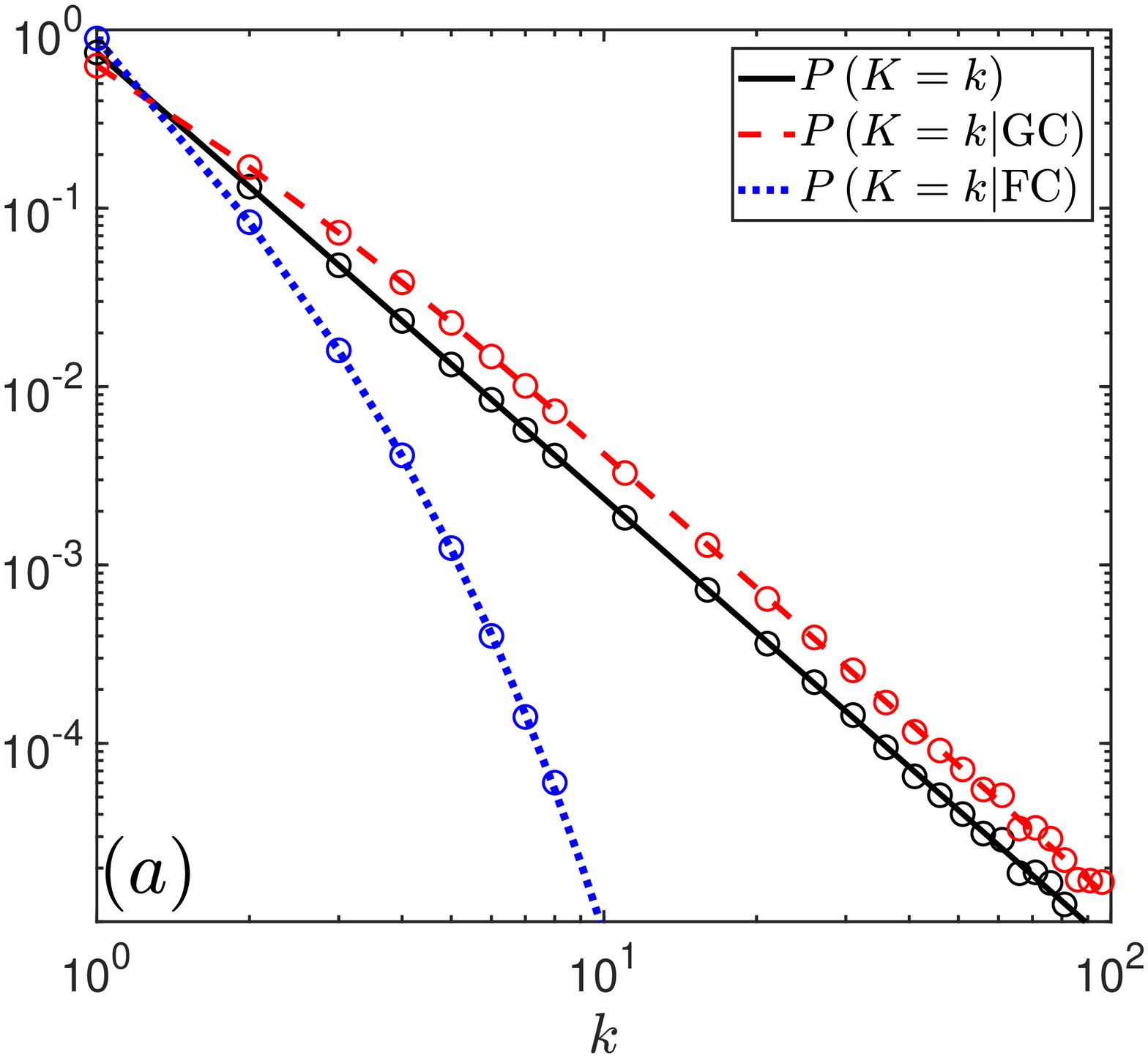}
%\\
\includegraphics[width=7cm]{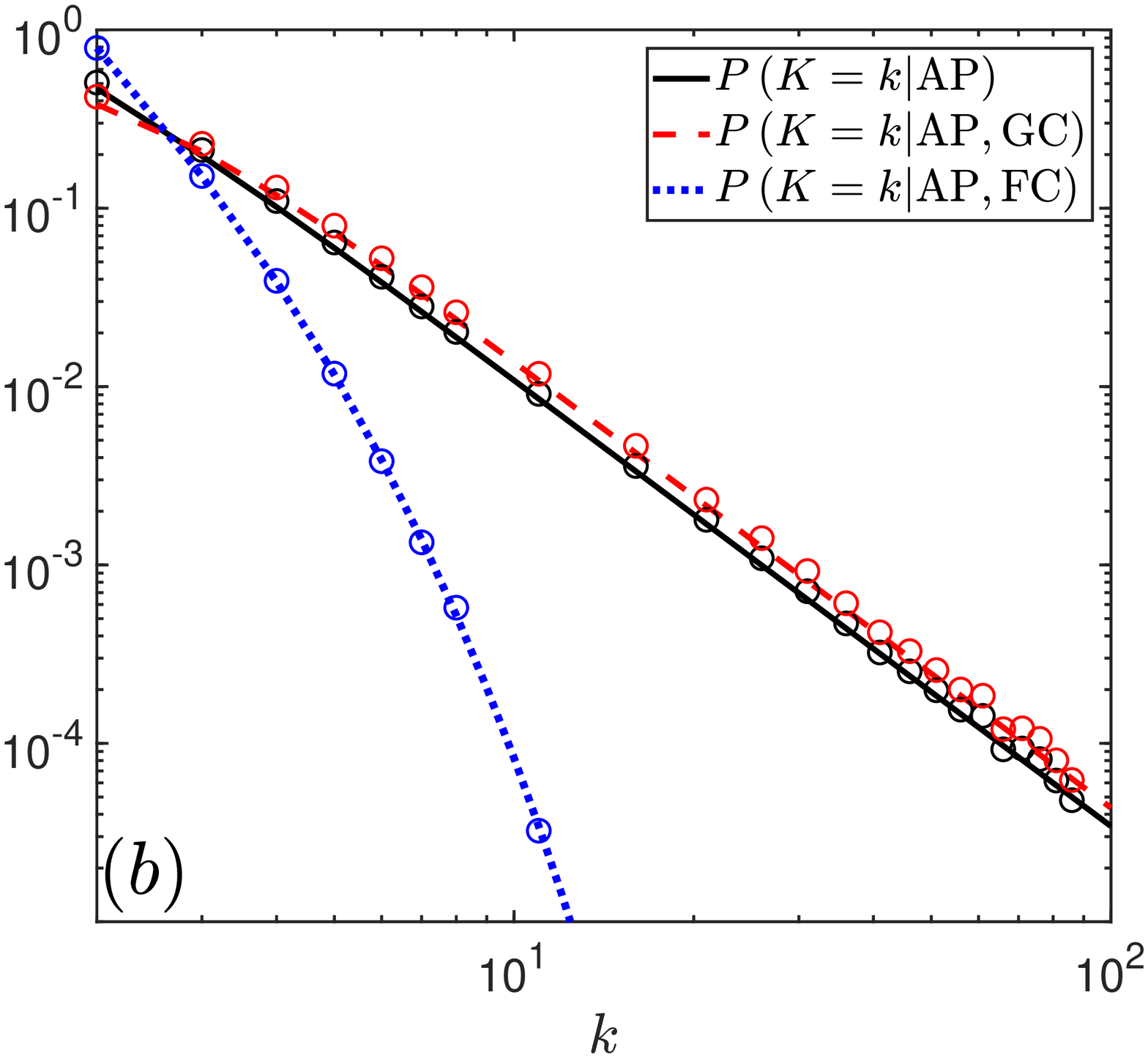}
\caption{
(Color online)
(a) Analytical results for the degree distribution
$P(K=k)$  
of a configuration model network 
with a power-law degree distribution (solid line),
where $k_{\rm min}=1$, $k_{\rm max}=100$,
$\gamma=2.5$ (and $c \simeq 1.8$),
the degree distribution
$P(K=k | {\rm GC})$ 
of the giant component (dashed line)
and the degree distribution
$P(K=k | {\rm FC})$ 
of the finite components (dotted line);
(b) Analytical results for the degree distribution
$P(K=k | {\rm AP})$  
of APs in a configuration model network with 
a power-law degree distribution (solid line),
where $k_{\rm min}=1$, $k_{\rm max}=100$,
$\gamma=2.5$ (and $c \simeq 1.8$),
the degree distribution
$P(K=k | {\rm AP},{\rm GC})$ 
of APs in the giant component (dashed line)
and the degree distribution
$P(K=k | {\rm AP},{\rm FC})$ 
of APs in the finite components (dotted line).
The analytical results are in very good agreement with the
results of computer simulations (circles).
}
\label{fig:10}
\end{figure}

\noindent
In Fig. \ref{fig:10}(a) 
we present
analytical results for the degree distribution
$P(K=k)$ (solid line) 
of a configuration model network 
with a power-law degree distribution,
where $k_{\rm min}=1$, $k_{\rm max}=100$ and $c=2$,
the degree distribution
$P(K=k | {\rm GC})$ (dashed line)
of the giant component
and the degree distribution
$P(K=k | {\rm FC})$ (dotted line)
of the finite components.
In Fig. \ref{fig:10}(b) 
we present analytical results for the degree distribution
$P(K=k | {\rm AP})$ (solid line) 
of APs in a configuration model network with 
a power-law degree distribution,
where $k_{\rm min}=1$, $k_{\rm max}=100$ and $c=2$,
the degree distribution
$P(K=k | {\rm AP},{\rm GC})$ (dashed line)
of APs in the giant component
and the degree distribution
$P(K=k | {\rm AP},{\rm FC})$ (dotted line)
of APs in the finite components.
The analytical results are in very good agreement with the
results of computer simulations.

\begin{figure}
\includegraphics[width=7cm]{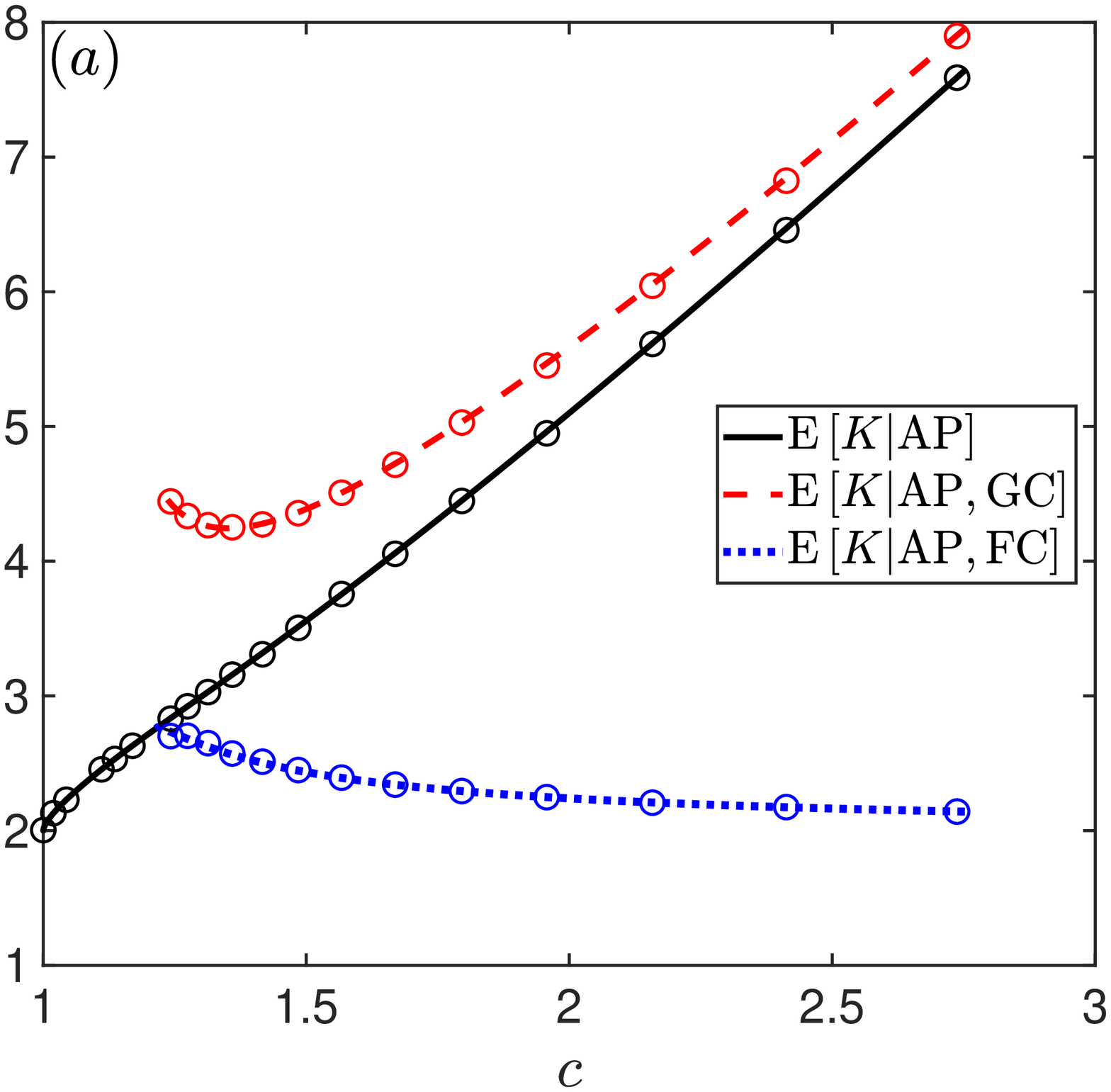}
%\\
\includegraphics[width=7cm]{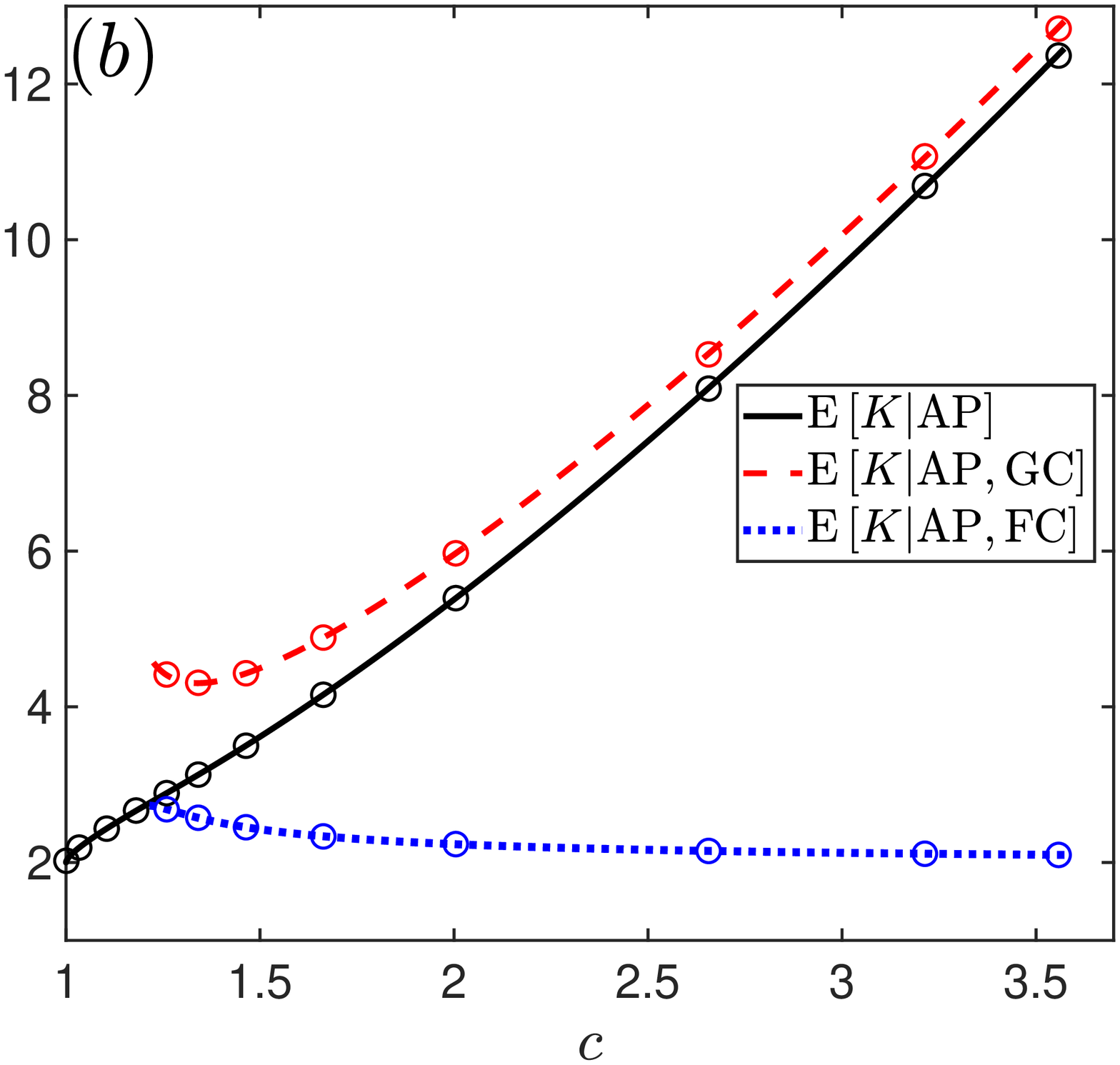}
\caption{
(Color online)
(a) Analytical results for the mean degree, 
$\mathbb{E}[K | {\rm AP}]$,
of APs 
as a function of $c$ in a
configuration model network 
with a power-law degree distribution (solid line),
where $k_{\rm min}=1$, $k_{\rm max}=100$,
the mean degree,
$\mathbb{E}[K | {\rm AP},{\rm GC}]$,
of APs in the giant component (dashed line)
and the mean degree,
$\mathbb{E}[K | {\rm AP},{\rm FC}]$,
of APs in the finite components (dotted line).
The analytical results 
are in very good agreement with the
results of computer simulations (circles)
for $N=4 \times 10^4$;
(b) Similar results for $k_{\rm max}=1000$ on a 
larger network of size $N=10^6$.
}
\label{fig:11}
\end{figure}

In Fig. \ref{fig:11} we present analytical results for
the mean degree of the APs 
$\mathbb{E}[K|{\rm AP}]$ (solid line),
in configuration model networks with a power-law degree distribution,
where $k_{\rm min}=1$ 
and $k_{\rm max}=100$ (a) or $k_{\rm max}=1000$ (b),
as a function of $c$,
We also present analytical results for the mean degree of APs which reside on the
giant component,
$\mathbb{E}[K|{\rm AP},{\rm GC}]$ (dashed line)
and the mean degree of APs which reside on the
finite components,
$\mathbb{E}[K|{\rm AP},{\rm FC}]$ (dotted line).
These results are in very good agreement with the results obtained
from computer simulations (circles).

The rank distribution is given by

\begin{equation}
P(R=r | {\rm GC}) =
\frac{\left(1-\tilde g \right)^r}{g[\zeta(\gamma) - \zeta(\gamma,k_{\rm max}+1)]}
\sum_{k=r+1}^{\infty}
\binom{k}{r} \tilde g^{k-r} k^{-\gamma}.
\label{eq:rAPL1SF}
\end{equation}

\begin{figure}
\includegraphics[width=7cm]{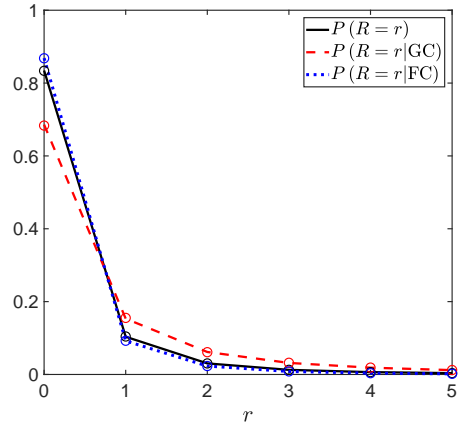}
\caption{
(Color online)
Analytical results for the distribution 
$P(R=r)$ 
of the ranks of nodes in a
configuration model network 
with a power-law degree distribution (solid line),
where $k_{\rm min}=1$, $k_{\rm max}=100$,
$\gamma=2.5$ and $c \simeq 1.8$.
The distribution
$P(R=r | {\rm GC})$ 
of the ranks of nodes in the giant component
(dashed line),
and the distribution
$P(R=r | {\rm FC})$ 
of the ranks of nodes in the finite components
(dotted line)
are also shown.
The analytical results
are in very good agreement with the
results of computer simulations (circles).
}
\label{fig:12}
\end{figure}

\noindent
In Fig. \ref{fig:12} we present analytical results for
the rank distribution
$P(R=r)$ (solid line)
of a configuration model network 
with a power-law degree distribution,
where $k_{\rm min}=1$, $k_{\rm max}=100$ and $\gamma=2.5$.
The distribution
$P(R=r | {\rm GC})$ (dashed line)
of the ranks of nodes in the giant component
and the distribution
$P(R=r | {\rm FC})$ (dotted line)
of the ranks of nodes in the finite components
are also shown.
The analytical results
are in very good agreement with the
results of computer simulations (circles).

\begin{figure}
\includegraphics[width=7cm]{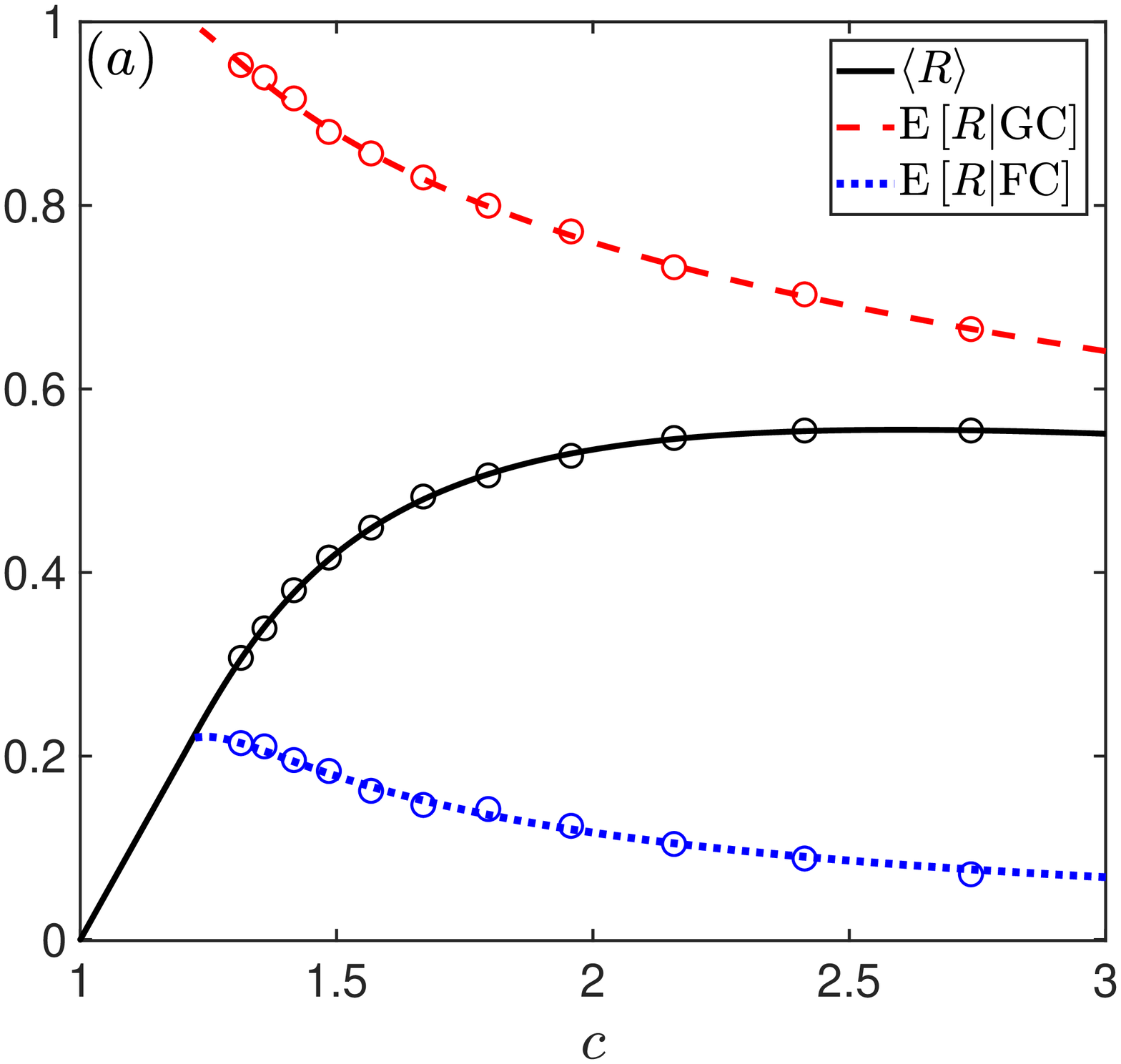}
%\\
\includegraphics[width=7cm]{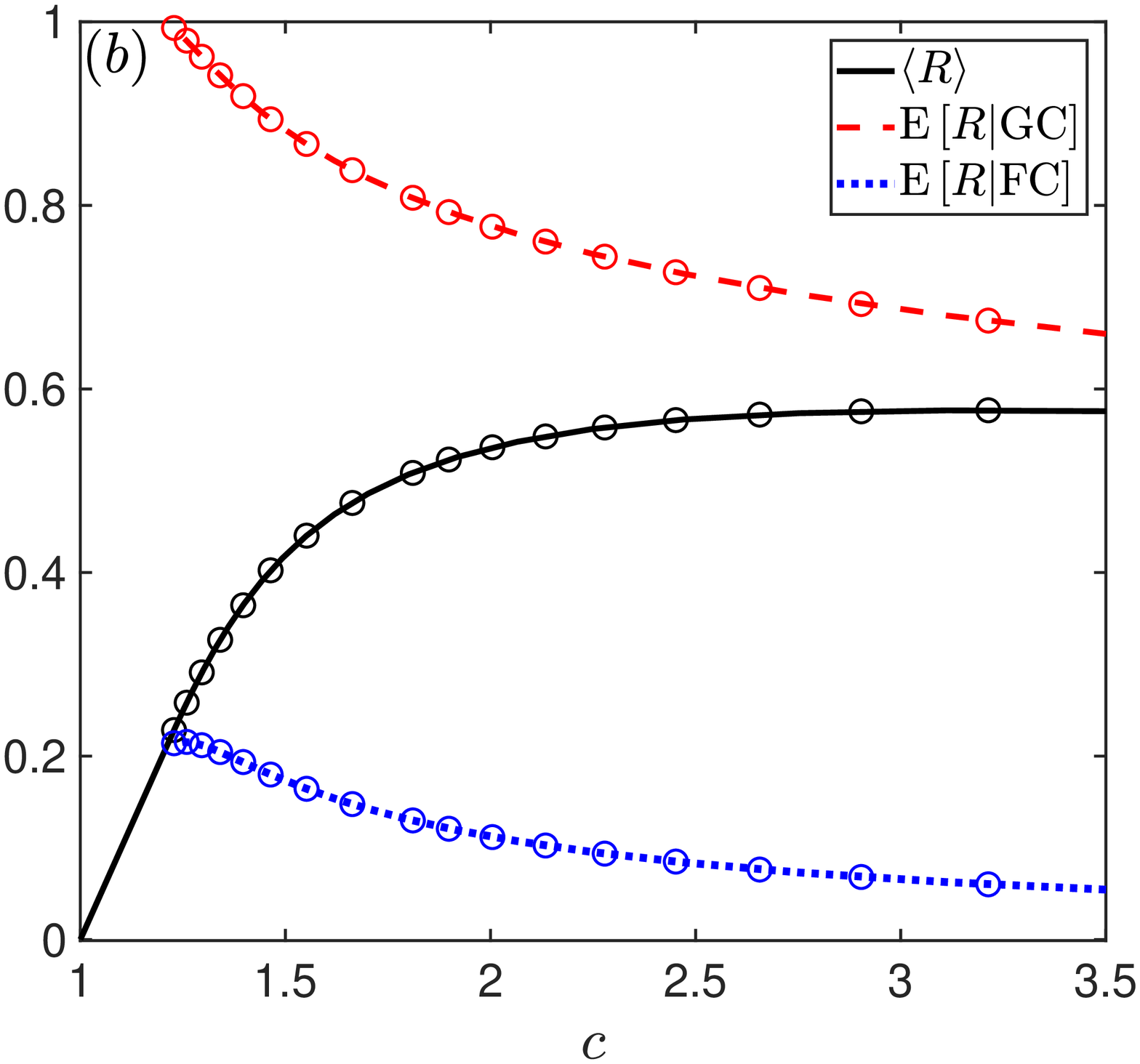}
\caption{
(Color online)
(a) Analytical results for the mean articulation rank,
$\langle R \rangle$,  
of a configuration model network 
with a power-law degree distribution (solid line)
with $k_{\rm max}=100$,
as a function of the mean degree, $c$.
The mean articulation rank
$\mathbb{E}[R|{\rm GC}]$ 
of the nodes which reside in the giant component
(dashed line)
and the mean articulation rank
$\mathbb{E}[R|{\rm FC}]$ 
of the nodes which reside in the finite components
(dotted line)
are also shown.
The analytical results 
are in very good agreement with the
results of computer simulations (circles)
for $N = 4 \times 10^4$;
(b) Similar results for $k_{\rm max}=1000$ on a 
larger network of size $N=10^6$.
}
\label{fig:13}
\end{figure}

In Fig. \ref{fig:13} we present
analytical results for the mean articulation rank,
$\langle R \rangle$,  
of a configuration model network with a power-law
degree distribution (solid line),
as a function of the mean degree, $c$,
the mean articulation rank
$\mathbb{E}[R|{\rm GC}]$ 
of the nodes which reside in the giant component
(dashed line)
and the mean articulation rank
$\mathbb{E}[R|{\rm FC}]$ 
of the nodes which reside in the finite components
(dotted line).
The analytical results 
are in very good agreement with the
results of computer simulations (circles).

Inserting the properties of configuration model networks with
a power-law degree distributions in Eq. (\ref{eq:Stub1}),
we find that the probability that a random $k=2$ AP that resides on the GC
is a tube is

\begin{equation}
P({\rm Tube}) = 1 
- \frac{ 1}{(1-\tilde g)[\zeta(\gamma-1) - \zeta(\gamma-1,k_{\rm max}+1) - 2^{1-\gamma} ]}.
\label{eq:TubeSF}
\end{equation}

\begin{figure}
\includegraphics[width=7cm]{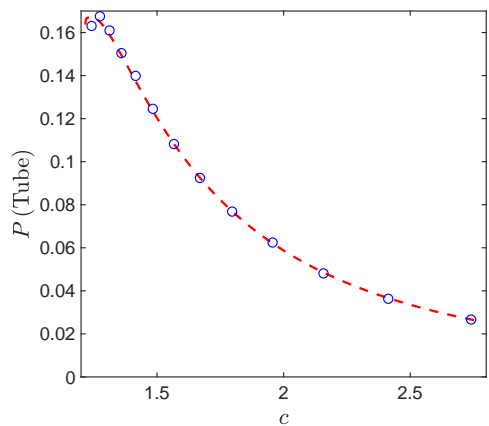}
\caption{
(Color online)
Analytical results (dashed line) for the
probability $P({\rm Tube})$ that a random $k=2$ AP on the GC 
of a configuration model network with a power-law degree distribution
is a tube,
as a function of the mean degree $c$,
obtained from Eq. (\ref{eq:TubeSF}).
The analytical results are in excellent agreement with the
results obtained from computer simulations (circles).
}
\label{fig:14}
\end{figure}

\noindent
In Fig. \ref{fig:14} we present the probability 
$P({\rm Tube})$ as a function of the
mean degree, $c$, for scale-free networks with $k_{\rm max}=100$. 
It is found that as the network becomes more dense
the trees that detach upon deletion of APs 
become of simpler topologies.

\section{Discussion}

Transportation, communication and many other networks consist of a single
connected component, in which there is at least one path connecting any
pair of nodes. This property is essential for the functionality of these networks.
Individual nodes in such networks may lose functionality due to inadvertent 
failures or intentional attacks.
The failure of a node disrupts the local processes taking place in that node
as well as the communication between this node and all the other nodes
in the network. In addition, such failure disconnects those paths
connecting other pairs of nodes, which go through that node.
In case that all the paths between nodes $j$ and $j'$ go through the
failed node, $i$, these two nodes end up in two disconnected components
of the network. In such case the failed node, $i$ is an AP, namely a node
whose deletion would break the network component on which it resides
into two or more components.
Networks which do not include any APs are called biconnected networks.
In such networks, each pair of nodes $j$ and $j'$ are connected to each
other by at least two non-overlapping paths.
These two non-overlapping paths form a closed loop, whose length
is equal to the sum of the lengths of the two paths.
Thus, in biconnected networks each pair of nodes resides on at least one
common cycle.
Thus, the deletion of any single node will leave at least one path between
any pair of other nodes intact.
While biconnected networks are resilient to the deletion of a single node,
they are still vulnerable to multiple node deletions. 
This is due to the fact that each node deletion may disconnect one of the
non-overlapping path between a pair of nodes $j$ and $j'$.
Moreover, the deletion of a node, $i$, often turns some other non-AP nodes into APs.
These newly formed APs are nodes that share a cycle with $i$, and do not
reside on any other cycle.
The replenishing of APs accelerates the dismantling of the network because
each time an AP is deleted it disconnects additional nodes from the networks.
The properties of APs are utilized in optimized algorithms of network dismantling
\cite{Braunstein2016,Zdeborova2016}.
The first stage of these dismantling processes is the decycling stage in
which one node is deleted in each cycle, transforming the network into a tree
network. In tree networks all the nodes of degrees $k \ge 2$ are APs and
thus the deletion of such nodes efficiently breaks the network into many
small components.

The sensitivity of a network to different types of attacks depends on properties
such as the degree distribution, correlations and the statistics of cycle lengths.
For example, it was shown that scale-free networks are resilient to random 
attacks but vulnerable to preferential attacks which target the high degree nodes
\cite{Cohen2001a}. 
Since high degree nodes are more likely to be
APs than low degree nodes, the deletion of a high degree
node is more likely to break the network into two or more components than
the deletion of a low degree node.

The role of APs in the resilience of complex networks to various random
and targeted attacks was recently studied using a combination of model
networks and a large set of empirical networks
\cite{Tian2017}.
Two AP-based attack scenarios were introduced.
In the AP-targeted attack, at each time step one deletes the most
destructive AP, namely the AP whose deletion removes the largest
number of nodes from the giant component.
It was found that in case that only a small fraction of the nodes are
deleted, this procedure leads to the fastest reduction in the size
of the giant component for a wide range of real-world networks.
In the greedy AP removal scenario, at each time step one 
simultaneously deletes all the APs that exist in the network at 
that time. Following the deletion new APs emerge, and the 
process continues until no new APs are formed and the 
remaining network becomes biconnected. The remaining
network is referred to as the residual giant bicomponent (RGB).
It was found that the fraction of nodes which are APs,
$P(i \in {\rm AP})$, 
and the
fraction of nodes which reside in the RGB,
$P(i \in {\rm RGB})$,
provide a useful characterization of the network.
To identify the topological characteristics that determine these
two quantities, the probabilities
$P(i \in {\rm AP})$
and 
$P(i \in {\rm RGB})$
were compared between each empirical network and its
randomized counterpart. 
Using a complete randomization which only maintains the number
of nodes, $N$ and the number of edges, $L$, it was found that
such randomization completely alters
$P(i \in {\rm AP})$
and 
$P(i \in {\rm RGB})$
and thus eliminates the topological characteristics that determine them.
In contrast, using degree-preserving randomization, which re-wires the
links while keeping the degree sequence unchanged, it was found that
these two quantities are not altered significantly
\cite{Tian2017}. 
This means that they are essentially encoded in the degree distribution
$P(K=k)$.
This implies that configuration model networks provide a
good description of the statistical properties of APs in complex
empirical networks with the same degree distribution.
Another useful property of the family of configuration model network
ensembles is that it is closed under random node deletion and preferential
node deletion processes. 
This means that when a configuration model network loses nodes
via random deletion or preferential deletion, its
degree distribution is modified accordingly, but it remains a 
configuration model network and does not form degree-degree
correlations.

\section{Summary}

We presented analytical results for the statistical properties
of articulation points in configuration model networks.
We obtained closed form expressions for the probability $P(i \in {\rm AP})$
that a random node in a network is an AP
and for the probability $P(i \in {\rm AP} | k)$
that a random node of a given degree, $k$, is an AP.
It is found that high degree nodes are more likely to be
APs than low degree nodes.
Using Bayes' theorem we obtained the degree distribution
$P(K=k | {\rm AP})$ of APs in the network.
It is found that APs of high degrees are more likely to reside
on the GC while APs of low degrees are more likely to reside on
the FCs.
Apart from its degree, each node can be characterized by its
articulation rank, $r$, which is the number of components
that would be added to the network upon deletion of that node.
Clearly, the articulation rank of a node of degree $k$ may take 
the values $r=0,1,2,\dots,k-1$.
For nodes which are not APs the articulation rank is $r=0$,
while the articulation ranks of APs satisfy $r \ge 1$.
We obtained a closed form expression for the distribution 
of articulation ranks,
$P(R=r)$.
To examine the distinct properties of APs on the giant and finite components,
we evaluated the probabilities presented above separately for the giant
and the finite components.
We applied these results to ensembles of configuration model networks
with degree distributions which follow 
a Poisson distribution (ER networks), 
an exponential distribution of the form $P(K=k) \sim e^{- \alpha k}$
where $k \ge k_{\rm min}=1$
and a power-law distribution 
of the form 
$P(K=k) \sim k^{- \gamma}$
(scale-free networks),
where $k_{\rm min} \le k \le k_{\rm max}$,
where $k_{\rm min}=1$.
It is found that for the Poisson and exponential degree distributions,
as the mean degree $c$ is increased, the fraction of APs in the network
increases in the sparse network regime, reaches a maximum value and then declines
as the dense network regime is approached.
In contrast, for the power-law distribution the behavior of
$P(i \in {\rm AP})$ depends on the value of $k_{\rm max}$.
Note that in scale-free networks the maximal value of $c$ is 
bounded from above by $k_{\rm max}$ according to Eq. (\ref{eq:cmax}).

\appendix

\section{Properties of the generating functions}

Using the properties of the generating functions $G_0(x)$ and $G_1(x)$,
one can obtain some inequalities between the values of $g$ and $\tilde g$.
The relation between $g$ and $\tilde g$ is expressed by Eq. (\ref{eq:g}),
which takes the form

\begin{equation}
1 - g = \sum_{k=0}^{\infty} (1-\tilde g)^k P(K=k).
\label{eq:1g}
\end{equation}

\noindent
For $0 < \tilde g < 1$, one can replace all the powers
$(1-\tilde g)^k$ with $k \ge 1$ by $1-\tilde g$
and obtain

\begin{equation}
1 - g < P(K=0) + \sum_{k=1}^{\infty} (1-\tilde g) P(K=k).
\end{equation}

\noindent
Expressing the sum on the right hand side in terms of $P(K=0)$
we obtain the inequality

\begin{equation}
1 - g < (1-\tilde g) + \tilde g P(K=0).
\end{equation}

\noindent
Therefore,

\begin{equation}
g > \tilde g [1-P(K=0)].
\label{eq:g>tg1mp}
\end{equation}

Similarly, by replacing all the powers
$(1-\tilde g)^k$ with $k \ge 2$ by $(1-\tilde g)^2$
we obtain a stronger constraint of the form

\begin{equation}
1 - g < (1-\tilde g)^2 + \tilde g (2-\tilde g) P(K=0) + \tilde g(1-\tilde g) P(K=1).
\end{equation}

\noindent
Expressing $g$ in terms of $\tilde g$, $P(K=0)$ and $P(K=1)$,
we obtain a lower bound for $g$, which is given by

\begin{equation}
g > \tilde g (2-\tilde g) - \tilde g (2-\tilde g) P(K=0) - \tilde g(1-\tilde g) P(K=1).
\label{eq:glb}
\end{equation}

To obtain an upper bound for $g$ we multiply
Eq. (\ref{eq:tg}) by $1-\tilde g$, and obtain

\begin{equation}
(1-\tilde g)^2 = 
\sum_{k=0}^{\infty} (1-\tilde g)^k  \widetilde P(K=k).
\label{eq:1tg2}
\end{equation}

\noindent
Comparing the right hand sides of Eqs. (\ref{eq:1g}) and (\ref{eq:1tg2}),
one observes that these equations provide the mean of $(1-\tilde g)^k$ under
$P(K=k)$ and $\widetilde P(K=k)$, respectively. 
Since $\widetilde P(K=k)$ gives more weight to large values of $k$,
while $(1-\tilde g)^k$ decreases monotonically as a function of $k$,
we conclude that for $0 < \tilde g < 1$

\begin{equation}
\frac{(1-\tilde g)^2}{1 - g} < 1.
\end{equation}

\noindent
Therefore, the upper bound of $g$ is given by

\begin{equation}
g < \tilde g (2-\tilde g).
\label{eq:gub}
\end{equation}

\noindent
In fact, this argument shows that for any $0 < x < 1$

\begin{equation}
x G_1(x) < G_0(x) - P(K=0).
\label{eq:xG1G0}
\end{equation}

\begin{figure}
\includegraphics[width=7.0cm]{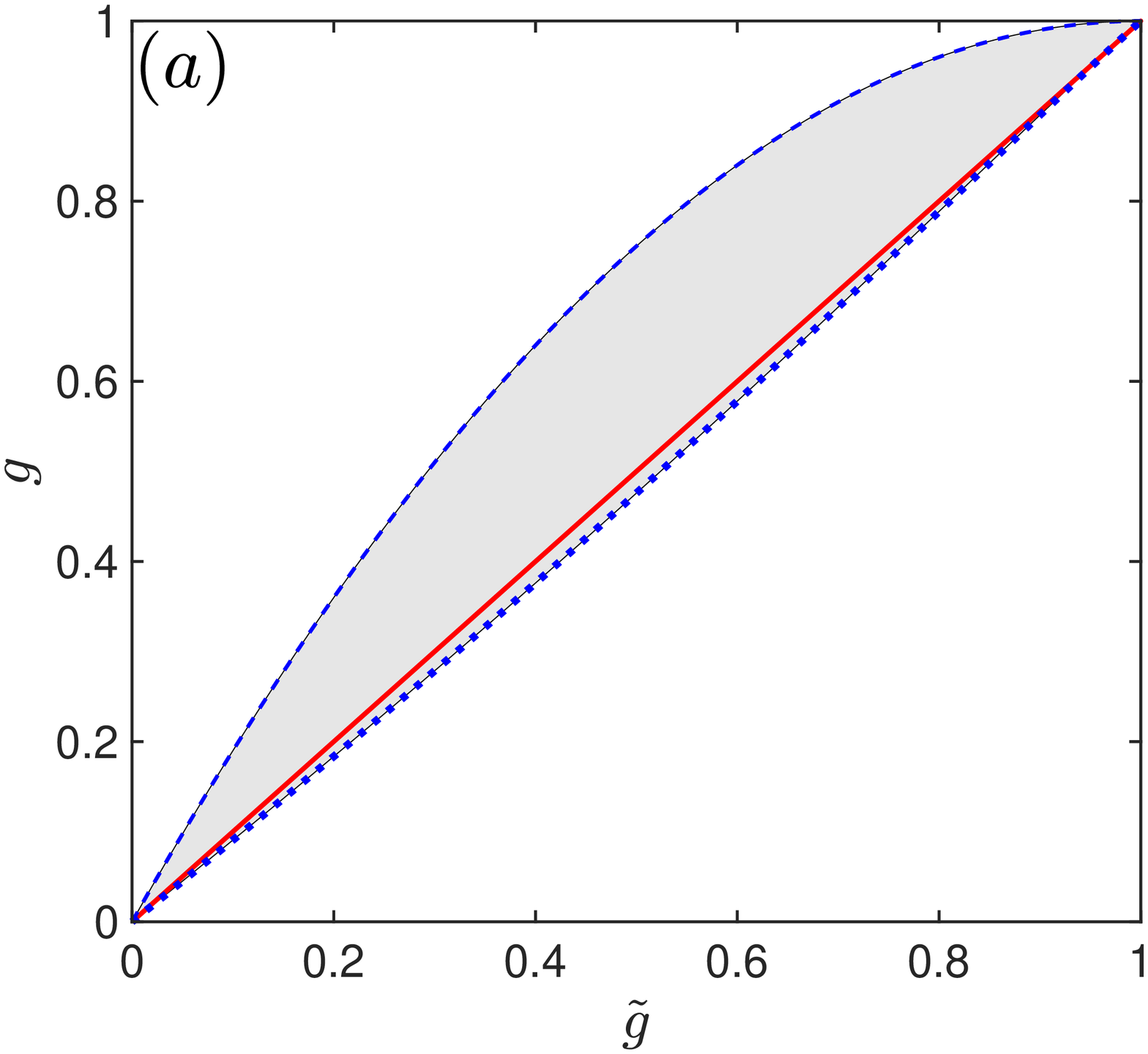}
%\\
\includegraphics[width=7.0cm]{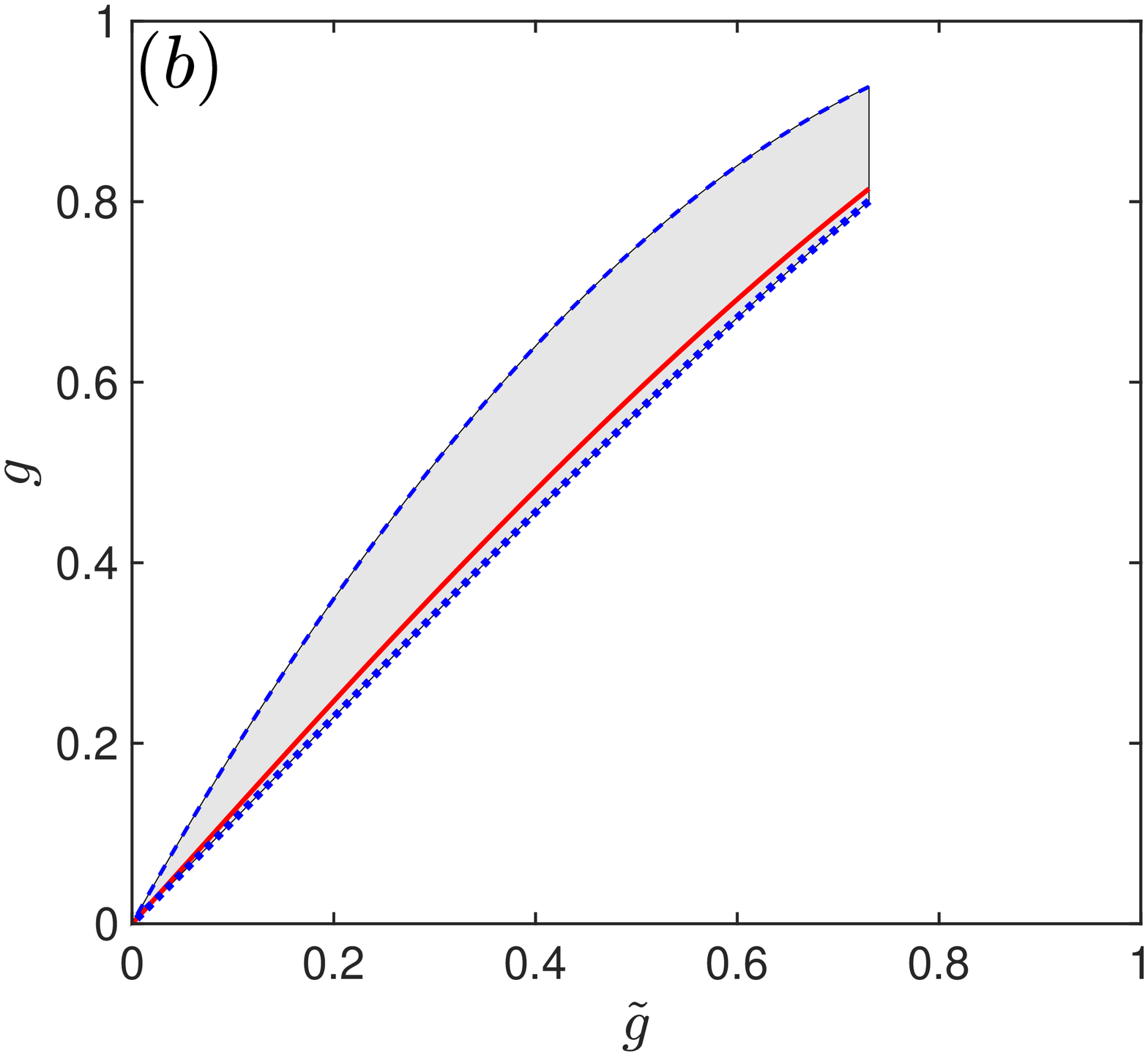}
\caption{
(Color online)
The probability $g$ that a random node resides in the giant component
as a function of the probability $\tilde g$ that a random neighbor of a random
node resides in the giant component for (a) ER networks and (b) configuration
model networks with a power-law degree distribution.
The shaded area shows the range of values of $g$ between the
lower bound (dotted line), given by Eq. (\ref{eq:glb}) and the
upper bound (dashed line), given by Eq. (\ref{eq:gub}).
In both networks the lower bound is very close
to the actual result, while the upper bound is significantly higher.
For the power-law degree distribution, 
it is found that for $k_{\rm max}=100$
the value of $g=g(\tilde g)$, obtained when the exponent
$\gamma$ is lowered towards $\gamma=2$ is around $0.8$.
In order to obtain larger values of $g$ one needs to further increase
the value of the upper cutoff, $k_{\rm max}$.
}
\label{fig:15}
\end{figure}

In Fig. \ref{fig:15} we show the
probability $g$ that a random node resides on the giant component (solid line)
as a function of the probability $\tilde g$ that a random neighbor of a random
node resides on the giant component for ER networks [Fig. \ref{fig:15}(a)]
and for configuration model networks with a power-law degree distribution
[Fig. \ref{fig:15}(b)].
The shaded area shows the range of values of $g$ between the
lower bound (dotted line), given by Eq. (\ref{eq:glb}) and the
upper bound (dashed line), given by Eq. (\ref{eq:gub}).
It is found that in both networks the lower bound is very close
to the actual result, while the upper bound is much higher.
In the case of the power-law degree distribution, 
it is found that for $k_{\rm max}=100$
the value of $g=g(\tilde g)$, obtained when the exponent
$\gamma$ is lowered towards $\gamma=2$ is around $0.8$.
In order to obtain larger values of $g$ one needs to increase
the value of the upper cutoff, $k_{\rm max}$.

Just above the percolation transition, where $\tilde g \ll 1$, one can expand
the term $(1-\tilde g)^{k-1}$ in Eq. (\ref{eq:tg}) to second order in $\tilde g$.
Solving the resulting equation, we obtain that in this limit

\begin{equation}
\tilde g = 2 \frac{ \langle K^2 \rangle - 2 \langle K \rangle }
{\langle K^3 \rangle - 3 \langle K^2 \rangle + 2 \langle K \rangle}.
\end{equation}

\noindent
Using a similar expansion for $(1-\tilde g)^k$ in Eq. (\ref{eq:g}), we find that
just above the percolation transition,  they are related by

\begin{equation}
g = \langle K \rangle \tilde g 
+ \frac{1}{2} \left( \langle K^2 \rangle - \langle K \rangle \right) \tilde g^2.
\end{equation}

\noindent
Thus, in the limit of $c \rightarrow {c_0}^{+}$, where $c_0$ is the mean degree
at the percolation transition, $0 < g,\tilde g \ll 1$ and

\begin{equation}
\lim_{c \rightarrow {c_0}^{+}}
\left( \frac{\tilde g}{g}  \right)
= \frac{\tilde g (1-\tilde g)}{g} 
= \frac{1}{\langle K \rangle}.
\label{eq:ggt0}
\end{equation}

\noindent
Combining this results with 
Eq. (\ref{eq:EkL1b})
it is found that just above the percolation transition,
the mean degree on the giant component 
is

\begin{equation}
\lim_{c \rightarrow {c_0}^{+}} \mathbb{E}[K | {\rm GC}] = 2.
\label{eq:EkL1bpc}
\end{equation}

\noindent
This result implies that unlike the value of the percolation
threshold $c_0$ which depends on the degree distribution,
the mean degree of the giant component upon its formation
is universally $\mathbb{E}[K | {\rm GC}] = 2$.

\section{The mean degree of the articulation points that reside on the giant component}

In this Appendix we show that the mean degree of APs on the giant component,
given by Eq. (\ref{eq:EkAPL1}),
is larger than the mean degree of all nodes on the giant component,
given by Eq. (\ref{eq:EkL1}).
To this end, we express Eq. (\ref{eq:EkAPL1}) in the form

\begin{equation}
\mathbb{E}[K|{\rm AP},{\rm GC}] =
\frac{ \tilde g (2-\tilde g) - \tilde g G_1(\tilde g) }
{g - [ G_0(\tilde g) - P(K=0) ]} \langle K \rangle.
\label{eq:EKAPL1A}
\end{equation}

\noindent
From Eq. (\ref{eq:xG1G0}) we obtain

\begin{equation}
G_0(\tilde g) - P(K=0) > \tilde g G_1(\tilde g).
\end{equation}

\noindent
Inserting this result into Eq. (\ref{eq:EKAPL1A}) we obtain

\begin{equation}
\mathbb{E}[K|{\rm AP},{\rm GC}] >
\frac{ \tilde g (2-\tilde g) - \tilde g G_1(\tilde g) }
{g - \tilde g G_1(\tilde g)} \langle K \rangle.
\label{eq:EKAPL1B}
\end{equation}

\noindent
From Eq. (\ref{eq:gub}) we know that

\begin{equation}
\frac{\tilde g (2-\tilde g)}{g} > 1.
\end{equation}

\noindent
Therefore,

\begin{equation}
\frac{\tilde g (2-\tilde g) - x}{g - x} > \frac{\tilde g (2-\tilde g)}{g}
\end{equation}

\noindent
for any value of $x$ in the range
$0 < x < \min \{ g, \tilde g(2-\tilde g) \}$.
The subtracted term $x = \tilde g G_1(\tilde g)$
clearly satisfies
$\tilde g G_1(\tilde g) < \tilde g (2-\tilde g)$.
This is due to the fact that $G_1(\tilde g) < 1$
while $2-\tilde g > 1$.
Since $\mathbb{E}[K|{\rm AP},{\rm GC}] > 0$,
one concludes that the condition
$\tilde g G_1(\tilde g) < g$
is also satisfied.
Therefore,

\begin{equation}
\frac{ \tilde g (2-\tilde g) - \tilde g G_1(\tilde g) }
{g - \tilde g G_1(\tilde g)} 
>
\frac{\tilde g(2-\tilde g)}{g}.
\end{equation}

\noindent
We thus conclude that

\begin{equation}
\mathbb{E}[K|{\rm AP},{\rm GC}] > \mathbb{E}[K|{\rm GC}].
\end{equation}

\section{Articulation points in the ternary network}

The statistical properties of APs in configuration model networks are
sensitive to the abundance of nodes of low degrees, particularly
nodes of degree $k=1$ (leaf nodes) and $k=2$.
In contrast, nodes of degree $k=0$ are excluded from the
giant component and are isolated from other finite components.
Thus, the isolated nodes can be discarded, with a suitable 
renormalization of the degree distribution.
In order to perform a systematic analysis of the statistical properties
of APs we consider a configuration model network with a ternary 
degree distribution of the form
\cite{Newman2010}

\begin{equation}
P(K=k) = p_1 \delta_{k,1} + p_2 \delta_{k,2} + p_3 \delta_{k,3},
\label{eq:ternary}
\end{equation}

\noindent
where $\delta_{k,n}$ is the Kronecker delta,
and 
$p_1+p_2+p_3=1$.
Since there are three non-zero probabilities and one
normalization condition, this model exhibits two free
parameters. It thus provides more flexibility than
the ER network and the configuration model networks
with an exponential and a power-law degree distribution, 
whose degree distributions are governed by a single parameter.
The mean degree of ternary network is given by

\begin{equation}
\langle K \rangle = p_1 + 2 p_2 + 3 p_3.
\label{eq:Ktern}
\end{equation}

\noindent
The generating functions are

\begin{equation}
G_0(x) = p_1 x + p_2 x^2 + p_3 x^3,
\label{eq:G0tern}
\end{equation}

\noindent
and

\begin{equation}
G_1(x) = \frac{ p_1  + 2 p_2 x + 3 p_3 x^2}{p_1 + 2 p_2 + 3 p_3}.
\label{eq:G1tern}
\end{equation}

\noindent
Solving Eq. (\ref{eq:tg}) for $\tilde g$, with $G_1(x)$ given by 
Eq. (\ref{eq:G1tern}), 
we find that

\begin{equation}
\tilde g = 
\begin{dcases}
0 &  \ \ \ \  p_3 \le \frac{p_1}{3} \\
1 - \frac{p_1}{3p_3} &  \ \ \ \  p_3 > \frac{p_1}{3}.
\end{dcases}
\label{eq:gtT}
\end{equation}

\noindent
Using Eq. (\ref{eq:g}) for $g$, where $G_0(x)$ is given by 
Eq. (\ref{eq:G0tern}),
we find that

\begin{equation}
g = 
\begin{dcases}
0 & \ \ \ \   p_3 \le \frac{p_1}{3} \\
1 - \left( \frac{p_1}{3p_3} \right) p_1 - \left( \frac{p_1}{3 p_3} \right)^2 p_2
- \left( \frac{p_1}{3 p_3} \right)^3 p_3 &
\ \ \ \   p_3 > \frac{p_1}{3}.
\end{dcases}
\label{eq:gT}
\end{equation}

\noindent
Thus, the percolation threshold is located at $p_3 = p_1/3$,
independently of $p_2$.
Using the normalization condition, we find that for any given
value of $p_2$, a giant component exists for 

\begin{equation}
p_3 > \frac{1-p_2}{4}.
\label{eq:p3gt}
\end{equation}

\noindent
For convenience we define the parameters

\begin{equation}
q_1 = \frac{p_1}{3 p_3}, 
\label{eq:q1} 
\end{equation}

\noindent
and

\begin{equation}
q_2 = \frac{p_2}{3 p_3}.
\label{eq:q2}  
\end{equation}

\noindent
From Eqs. (\ref{eq:gtT})  and (\ref{eq:gT}), one can see that
a necessary and sufficient condition for the existence of a
giant component is that $q_1$ will be in the range $0 \le q_1 < 1$.
In contrast, there is no such condition on $q_2$, namely a giant
component may exist for any value of $q_2 \ge 0$.

The degree distribution of the giant component is given by

\begin{equation}
P(K=k | {\rm GC}) = \frac{ 1 - q_1^k }
{1 - q_1 p_1
- q_1^2 p_2
- q_1^3 p_3 } 
P(K=k),
\label{eq:Pk1tern}
\end{equation}

\noindent
where $k=1, 2, 3$ and $P(k)$ is given by Eq. (\ref{eq:ternary}).
The degree distribution on the finite components is given by

\begin{equation}
P(K=k | {\rm FC}) = \frac{ q_1^k }
{ q_1 p_1
+ q_1^2 p_2
+ q_1^3 p_3 } 
P(K=k).
\label{eq:Pk0tern}
\end{equation}

\noindent
Thus, the mean degree on the giant component is given by

\begin{equation}
\mathbb{E}[K | {\rm GC}] = \frac{ q_1^2}
{1 - q_1 p_1
- q_1^2 p_2
- q_1^3 p_3 } 
\langle K \rangle,
\label{eq:Ek1tern}
\end{equation}

\noindent
while the mean degree on the finite components is given by

\begin{equation}
\mathbb{E}[K | {\rm FC}] = \frac{ q_1^2 }
{ q_1 p_1
+ q_1^2 p_2
+ q_1^3 p_3 } 
\langle K \rangle.
\label{eq:Ek0tern}
\end{equation}

\noindent
Using Eq. (\ref{eq:AP2}) we obtain the probability that a random node is an
AP, which is given by

\begin{equation}
P(i \in {\rm AP}) = q_1 \left[ (2-q_1)(p_2+p_3) + (1-q_1)^2 p_3 \right].
\label{eq:APt}
\end{equation}

\noindent
From Eq. (\ref{eq:APL1}) we obtain the probability that a random node which
resides on the giant component is an AP, which is

\begin{equation}
P(i \in {\rm AP}| {\rm GC}) = \frac{3 q_1 (2 q_2 +1)}{1+4 q_1 + 3 q_2 + q_1^2 + 3 q_1 q_2},
\label{eq:APt1}
\end{equation}

\noindent
while from Eq. (\ref{eq:APL0}) we obtain the probability that a random node which
resides on one of the finite components is given by

\begin{equation}
P(i \in {\rm AP}| {\rm FC}) = \frac{ q_1 + 3 q_2}{3 + q_1 + 3 q_2}.
\label{eq:APt0}
\end{equation}

\noindent
An interesting question that arises is whether APs are more abundant on the giant
component or on the finite components, namely whether
$P(i \in {\rm AP}| {\rm GC})$ 
is larger or smaller than
$P(i \in {\rm AP}| {\rm FC})$.
On the one hand, the giant component includes nodes of higher degrees,
which are more likely to be APs and a smaller fraction of leaf nodes, which
cannot be APs. On the other hand, in the finite components
all the nodes of degrees $k \ge 2$ are APs, unlike the giant component which
exhibit cycles which reduce the number of APs.
From inspection of Eqs. (\ref{eq:APt1}) and (\ref{eq:APt0}) we conclude that
in the limit of $q_1 \rightarrow 1$, one obtains
$P(i \in {\rm AP}|{\rm GC}) > P(i \in {\rm AP}|{\rm FC})$.
In contrast, in the limit of $q_1 \ll 1$ we obtain
$P(i \in {\rm AP}|{\rm GC}) < P(i \in {\rm AP}|{\rm FC})$.

\begin{figure}
\includegraphics[width=7.0cm]{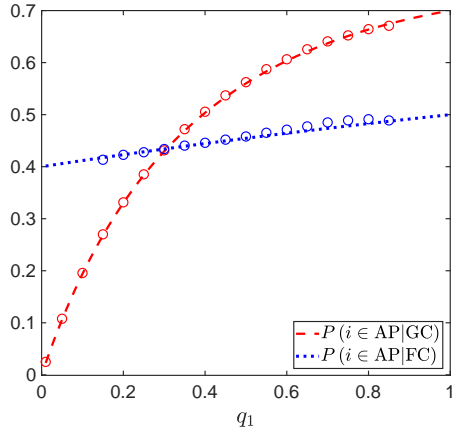}
\caption{
(Color online)
The probability $P(i \in {\rm AP}|{\rm GC})$ that a random node in the giant
component of a configuration model network with a ternary degree distribution
is an AP (dashed line) and the probability $P(i \in {\rm AP}|{\rm FC})$ that
a random node in one of the finite components is an AP (dotted line),
as a function of $q_1$, for $q_2=0.6$. It is found that for small values of $q_1$
the two probabilities satisfy
$P(i \in {\rm AP}|{\rm GC}) < P(i \in {\rm AP}|{\rm FC})$,
while for large values of $q_1$ they satisfy
$P(i \in {\rm AP}|{\rm GC}) > P(i \in {\rm AP}|{\rm FC})$.
It implies that when leaf nodes are scarce, APs are more abundant on
the finite components, and when leaf nodes are abundant, APs are more
more abundant on the giant component.
}
\label{fig:16}
\end{figure}

In Fig. \ref{fig:16} we present the
probability $P(i \in {\rm AP}|{\rm GC})$ that a random node in the giant
component of a configuration model network with a ternary degree distribution
is an AP (dashed line) and the probability $P(i \in {\rm AP}|{\rm FC})$ that
a random node in one of the finite components is an AP (dotted line),
as a function of $q_1$, for $q_2=0.6$. It is found that for small values of $q_1$
the two probabilities satisfy
$P(i \in {\rm AP}|{\rm GC}) < P(i \in {\rm AP}|{\rm FC})$,
while for large values of $q_1$
$P(i \in {\rm AP}|{\rm GC}) > P(i \in {\rm AP}|{\rm FC})$.
This implies that the relative abundances of APs in the giant
component and in the finite components depend on the 
parameters of the network.

The probability that a randomly selected AP resides on the giant component
is given by

\begin{equation}
P(i \in {\rm GC}|{\rm AP}) = \frac{ 3 (1-q_1)(2 q_2 + 1)}{(2-q_1)(3 q_2+1) + (1-q_1)^2},
\label{eq:L1APtern}
\end{equation}

\noindent
while the probability that a randomly selected AP resides on one of the
finite components is given by

\begin{equation}
P(i \in {\rm FC}|{\rm AP}) = \frac{q_1(q_1+3 q_2)}{(2-q_1)(3 q_2+1) + (1-q_1)^2}.
\label{eq:L0APtern}
\end{equation}

\noindent
From inspection of Eqs. (\ref{eq:L1APtern}) and (\ref{eq:L0APtern}) we find that in
the limit of $q_2 \gg 1$ and $q_1 \ll 1$ the probability that a randomly
selected AP resides on the giant component satisfies
$P(i \in {\rm GC}|{\rm AP}) \rightarrow 1$, 
while the probability that such
node resides on one of the finite components vanishes.

\end{document}